\documentclass[12pt]{article}
\usepackage{bbm}
\usepackage{bm}
\usepackage{textgreek}
\usepackage[bbgreekl]{mathbbol}
\usepackage{color}
\usepackage{authblk}
\usepackage{latexsym}
\usepackage{epsfig,amssymb,euscript, mathrsfs,cite}
\usepackage{amsmath}
\usepackage{bbold}
\usepackage{tikz}
\usetikzlibrary{decorations.pathmorphing}
\usepackage{slashed}
\usepackage{mathrsfs}
\usepackage{enumerate}
\usepackage{graphicx}
\usepackage{subcaption}
\definecolor{MyDarkBlue}{rgb}{0.15,0.15,0.45}
\usepackage[linktocpage=true]{hyperref}
\hypersetup{
colorlinks=true,
citecolor=MyDarkBlue,
linkcolor=MyDarkBlue,
urlcolor=MyDarkBlue,
pdfauthor={},
pdftitle={},
pdfsubject={hep-th}
}
\usepackage[toc,page]{appendix}

\newcommand{\nocontentsline}[3]{}
\newcommand{\tocless}[2]{\bgroup\let\addcontentsline=\nocontentsline#1{#2}\egroup}

\newsavebox{\ns}
\newsavebox{\dbrane}
\newsavebox{\dbshort}

\def\be{\begin{equation}}
\def\ee{\end{equation}}
\def\bea{\begin{eqnarray}}
\def\eea{\end{eqnarray}}

\newcommand{\nn}{\nonumber\\}

\newcommand\R{\mathbb{R}}
\newcommand\Z{\mathbb{Z}}

\newcommand\diff{\mathrm{d}}
\newcommand{\de}{\partial}

\newcommand{\dd}{\mathrm{d}}

\newcommand{\ii}{\mathrm{i}}

\newcommand{\ex}{\mathrm{e}}
\newcommand{\vol}{\mathrm{vol}}

\newcommand{\ssigma}{\sigma}

\newlength{\sswidth}

\def\rme{\mathrm{e}}

\numberwithin{equation}{section}       

\newcommand{\gammathree}{\mbox{\textgamma}}

\newcommand{\bk}{\kappa}
\newcommand{\SSigma}{\mathbb{\Sigma}}
\newcommand{\SBH}{S_{\mathrm{BH}}}
\newcommand{\nullH}{V}
\newcommand{\bb}{b}
\newcommand{\bc}{c}
\newcommand{\bt}{s}
\newcommand{\bR}{\rho}

\newcommand{\sps}{\mathfrak{s}}

\begin{document}

\begin{titlepage}

\begin{flushright}
Imperial/TP/2021/JG/04
\end{flushright}

\vskip 1cm

\begin{center}


\vskip 1cm

{\Large \bf Thermodynamics of accelerating and\\[2mm] supersymmetric $AdS_4$ black holes}

\vskip 1cm

\vskip 1cm
{Davide Cassani$^{\mathrm{a}}$, Jerome P. Gauntlett$^{\mathrm{b}}$, Dario Martelli$^{\mathrm{c,d,e}}$ and James Sparks$^{\mathrm{f}}$}

\vskip 1cm

${}^{\mathrm{a}}$\textit{INFN, Sezione di Padova,  Via Marzolo 8, 35131 Padova, Italy\\}

\vskip 0.2cm

${}^{\mathrm{b}}$\textit{Blackett Laboratory, Imperial College, \\
Prince Consort Rd., London, SW7 2AZ, U.K.\\}

\vskip 0.2cm

${}^{\mathrm{c}}$\textit{Dipartimento di Matematica ``Giuseppe Peano'', Universit\`a di Torino,\\
Via Carlo Alberto 10, 10123 Torino, Italy}

\vskip 0.2cm

${}^{\mathrm{d}}$\textit{INFN, Sezione di Torino \&}   ${}^{\mathrm{e}}$\textit{Arnold--Regge Center,\\
 Via Pietro Giuria 1, 10125 Torino, Italy}

\vskip 0.2cm

${}^{\mathrm{f}}$\textit{Mathematical Institute, University of Oxford,\\
Andrew Wiles Building, Radcliffe Observatory Quarter,\\
Woodstock Road, Oxford, OX2 6GG, U.K.\\}

\vskip 0.2 cm

\end{center}

\vskip 0.5 cm

\begin{abstract}
\noindent
We study the thermodynamics of $AdS_4$ black hole solutions of Einstein-Maxwell theory that are accelerating, rotating, and carry electric and magnetic charges. We focus on the class for which the black hole horizon is a spindle 
and can be uplifted on regular Sasaki-Einstein spaces to give solutions 
of $D=11$ supergravity that are free from conical singularities.  We use holography to calculate the Euclidean on-shell action and to define a set of conserved charges which give rise to a first law. 
We identify a complex locus of supersymmetric and non-extremal solutions, defined through an analytic continuation of the parameters, upon which we obtain a simple expression for the on-shell action. A Legendre transform of this action combined with a 
reality constraint then leads to the Bekenstein-Hawking entropy for the class of supersymmetric and extremal black holes.

\
\end{abstract}

\end{titlepage}

\pagestyle{plain}
\setcounter{page}{1}
\newcounter{bean}
\baselineskip18pt

\tableofcontents

\newpage
\section{Introduction}

The study of black hole thermodynamics in the context of the AdS/CFT correspondence continues to be a very active area of research. Focusing on the class of supersymmetric black holes in $AdS$ spacetime with dimension $D>3$, there has been significant progress in quantitatively recovering the Bekenstein-Hawking entropy by analysing 
appropriate statistical ensembles of the dual superconformal field theory (SCFT), starting with \cite{Benini:2015eyy,Benini:2016rke} for $D=4$ and \cite{Cabo-Bizet:2018ehj,Choi:2018hmj,Benini:2018ywd} for $D=5$.
In this context, the problem of microstate counting via holography is conveniently reformulated in terms of a supersymmetric field theory path integral in a background with sources. This takes the form of a supersymmetric index that 
can then be computed using a variety of methods and then compared with the black hole entropy; for example in the case of
$D=4$, which is the focus of this paper, see
\cite{Benini:2015noa,Hosseini:2016tor,Benini:2016hjo,Cabo-Bizet:2017jsl,Hosseini:2017fjo,Benini:2017oxt,Bobev:2018uxk,Gang:2019uay,Choi:2019zpz,Nian:2019pxj,Bobev:2019zmz,Benini:2019dyp,Hosseini:2019and}. On the gravity side, the same partition function is obtained from a suitably defined on-shell action, regularized so that the supersymmetric and extremal limit is well-defined, see e.g.\cite{Halmagyi:2017hmw,Azzurli:2017kxo,Cabo-Bizet:2017xdr,Bobev:2020pjk} for static black holes in $AdS_4$ and \cite{Cabo-Bizet:2018ehj,Cassani:2019mms} for the rotating case, where it was necessary to identify a novel complex locus of supersymmetric solutions.

In this paper we study various aspects of the thermodynamics of a class of $AdS_4$ black holes with non-zero acceleration.  The black holes are solutions of $D=4$ minimal gauged supergravity, whose bosonic content is simply Einstein-Maxwell theory with a negative cosmological constant. The solutions of interest lie within the 
Pleba\'nski-Demia\'nski (PD) family of solutions \cite{Plebanski:1976gy} and are also rotating 
as well as carrying both electric and magnetic charge. 
A consequence of the acceleration is that the black hole event horizon has conical singularities which stretch all the way out to the $AdS_4$ boundary. From a physical point of view these singularities can be interpreted as the tensions of 
``cosmic strings'' which pierce the horizon and give rise to the acceleration. 

Remarkably, these conical singularities can be completely removed after 
imposing suitable restrictions on the physical parameters and then embedding the solutions into $D=11$ supergravity \cite{Ferrero:2020twa}. 
Recall that any solution of $D=4$ minimal gauged supergravity can be uplifted on a seven-dimensional Sasaki-Einstein space to obtain a local solution of $D=11$ supergravity. 
One key ingredient in the construction of \cite{Ferrero:2020twa} is that
the Sasaki-Einstein manifold should be taken to lie in the regular class, meaning it is a $U(1)$ fibration over a six-dimensional K\"ahler-Einstein base; the simplest examples are $S^7$, as well as the discrete quotients $S^7/\mathbb{Z}_2$, $S^7/\mathbb{Z}_4$, all of which which are $U(1)$ fibrations over $\mathbb{CP}^3$. A second key ingredient is to impose suitable restrictions on the parameters of the $D=4$ black hole solution and, in particular, 
the conical deficits are assumed to be quantized 
so that the black hole horizon is a ``spindle''. Topologically
the spindle is a two-sphere but there are orbifold singularities
at both poles specified by two integers $n_\pm\in\mathbb{N}$ or, in other words, the spindle is
the weighted projective space $\mathbb{WCP}^1_{[n_-,n_+]}$. It is also the case that the magnetic charge of the black hole   is fixed by the spindle data and given by $G_{(4)}Q_m=(n_--n_+)/(4n_-n_+)$ and is always non-vanishing since we must have $n_-\ne n_+$.
By combining these ingredients, it was shown in \cite{Ferrero:2020twa} that for a spindle with given $n_\pm$, 
together with a suitably chosen regular Sasaki-Einstein manifold, the $D=11$ solution is free from any conical singularities.\footnote{The construction is essentially the same as that used for a class of $AdS_3\times\mathbb{WCP}^1_{[n_-,n_+]}$ solutions of minimal $D=5$ gauged supergravity which are uplifted on $SE_5$ to obtain solutions
of type IIB supergravity and describe D3-branes wrapping a spindle \cite{Ferrero:2020laf}. Related work on spindles appears in
\cite{Hosseini:2021fge,Boido:2021szx,Bah:2021mzw,Ferrero:2021wvk,Bah:2021hei}.}
Furthermore, the $D=11$ solutions preserve supersymmetry when 
the $D=4$ solutions do.

While we are principally interested in this regular class of black hole solutions in $D=11$, we will analyse their thermodynamics
in $D=4$ and the details of the internal Sasaki-Einstein space will not be important. The regular $D=4$ black hole solutions, by which we mean the $D=4$ solutions which can be uplifted to give regular $D=11$ solutions, 
can be specified by three physical parameters, the mass, $M$, angular momentum, $J$, and electric charge $Q_e$, along with the spindle data, $n_\pm$.  
For a fixed spindle, by varying the parameters appearing in the solution,
we will show that this three-parameter family of black holes satisfies a first law which takes the standard form.
An interesting feature is that the conformal boundary of these black hole solutions is not conformally flat.
  Our first law involves varying the conformal class of the boundary and also involves a specific rescaling of the time 
  coordinate\footnote{It would be interesting to see how our derivation of the first law fits into the approach of \cite{Papadimitriou:2005ii}.}, 
  as also seen in the derivation of the first law for accelerating black holes with vanishing magnetic charge given in \cite{Anabalon:2018qfv}. By calculating the Euclidean on-shell action we also derive a quantum statistical relation for the dual CFT at finite temperature.

The three-parameter family of regular $D=4$ black hole solutions includes a locus of supersymmetric solutions which satisfy the BPS relation
\begin{align}\label{bpslocusintro}
M\, = \, \frac{2}{\chi}J+Q_e\,,
\end{align}
where $\chi=(n_++n_-)/(n_+n_-)$ is the Euler character of the spindle.  
Our analysis will assume non-zero acceleration throughout, but many of our main results can be seen
as generalizations of the well-studied electrically charged, $AdS$ Kerr-Newman black holes with spherical horizons 
\cite{Carter:1968ks,Kostelecky:1995ei,Caldarelli:1998hg}, provided that we \emph{formally} set $n_-=n_+=1$, so that $\chi=2$, the Euler character of the two-sphere, and $Q_m=0$. For example, the relation \eqref{bpslocusintro} is then identical to that of the Kerr-Newman black holes. For the Kerr-Newman black holes
it is well known that  the condition for supersymmetry does not coincide with the condition that the black holes are extremal
and the same is true for the accelerating black holes. The regular, supersymmetric and extremal accelerating black holes are 
specified by a single parameter which can be taken to be the electric charge $Q_e$. The angular momentum $J$ 
and Bekenstein-Hawking entropy, $S_{BH}$, are then given by \cite{Ferrero:2020twa}, 
\begin{align}\label{extafterlegintro}
J \, = \,   \frac{Q_e}{4}\left(-\chi+\sqrt{\chi^2+(4G_{(4)})^2 \left(Q_e^2+Q_m^2\right)}\right)\, , \qquad \SBH\, = \, \frac{\pi}{G_{(4)}}\frac{J}{Q_e}\,, 
\end{align}
Notice that these expressions again formally reduce to the corresponding relation
for the supersymmetric and extremal Kerr-Newman black holes.

In this paper we will focus on black holes with non-vanishing rotation, $J\ne 0$, with the associated supersymmetric and extremal black holes then having $Q_e\ne 0$. This is partly because the case with $J=0$ was analysed in some detail in \cite{Ferrero:2020twa} and also because the $J=0$ solutions have some specific features which are not present when $J\ne 0$. For example, 
with $J=0$ the supersymmetric and extremal black holes, which arise when $Q_e=M=0$,
have an acceleration horizon that splits the conformal boundary into two halves and, furthermore, supersymmetry of the boundary is then preserved as a result of two different topological twists, one on each half. By contrast when $J\ne 0$ the conformal boundary is always regular and the boundary supersymmetry is not preserved as a result of a topological twist. In fact, here we will clarify how supersymmetry is preserved on the boundary when $J\ne 0$ by explicitly constructing the boundary spinors that solve the conformal Killing spinor equation, as expected on general grounds \cite{Hristov:2013spa}.

At finite temperature the Euclidean on-shell action, $I(T,\Phi_e,\Omega)$, of the black hole solutions can be identified with minus the logarithm of the partition function of the dual field theory. The action is a function of the temperature $T$ and the electric and rotational chemical potentials, 
$\Phi_e$ and $\Omega$, respectively, and the partition function is in a grand canonical ensemble. At finite temperature we can obtain the entropy of the black holes as a function of the mass $M$ and charges $J,Q_e$ in a microcanonical ensemble via a Legendre transformation using the quantum statistical relation. For extremal black holes at zero temperature this prescription breaks down because the on-shell action becomes ill-defined.  However, motivated by recent progress for the Kerr-Newman and other black holes \cite{Cabo-Bizet:2018ehj,Cassani:2019mms}, we can develop an analogous prescription for supersymmetric and extremal black holes.

 To do this we first introduce a complex locus of supersymmetric solutions that is obtained by 
 an analytic continuation of some of the parameters appearing in the black hole solutions. The 
one-parameter family of supersymmetric and extremal black holes, which are of course real, are then obtained 
on a special slice of
 this complex locus. By analysing various thermodynamic quantities, analytically continued to this complex locus, we are able to derive an expression for the on-shell Euclidean 
action $I=I(\omega,\varphi)$, expressed as
 a function of rotational and electric chemical potentials, $\omega$ and  $\varphi$, respectively, both of which are complex and defined on the supersymmetric locus. 
   Specifically, we find that the on-shell action can be written in the form
  \begin{align}\label{SUSYactionintro}
I (\omega,\varphi)&\, = \, \pm \frac{1}{2\ii G_{(4)}}\left[\frac{\varphi^2}{\omega} +  (G_{(4)}Q_m)^2\omega \right]\, ,
\end{align}
 with $\omega$ and  $\varphi$ satisfying the constraint
 \begin{align}\label{SUSYconintro}
\varphi - \frac{\chi}{4}\omega \, = \, \pm \ii\pi\,.
\end{align}
By carrying out a Legendre transform, or more precisely extremizing the quantity, $- I(\omega, \varphi) - \omega J - \varphi Q_e$, which is sometimes called an entropy function, subject to the constraint \eqref{SUSYconintro}, we obtain
an expression for the entropy $\SBH(J,Q_e)$ and charges $J,Q_e$.
By imposing the condition that $\SBH$, $Q_e$ and $J$ are all real, we then precisely recover the 
expressions for the entropy and angular momentum for the supersymmetric and extremal black holes given in \eqref{extafterlegintro}. 
Our result extends the extremization principles that have been
formulated for $D=4$ rotating and dyonically charged black holes in ~\cite{Choi:2018fdc,Hosseini:2019iad} (see also~\cite{Hosseini:2017mds,Hosseini:2018dob} for earlier results in different dimensions) and in this context it is worth highlighting again that our magnetically charged black holes are not preserving supersymmetry via a topological twist.

In a separate calculation, we show that the expression \eqref{SUSYactionintro} for the on-shell action defined on the complex locus of solutions can also be obtained from a suitable analytic continuation of
the result of \cite{BenettiGenolini:2019jdz}, which calculates the action using the fixed point data of the supersymmetric Killing vector obtained as a bi-linear of the bulk Killing spinors.
 It would be of much interest to derive our result for the on-shell action, $I=I(\omega,\varphi)$, directly from the dual SCFT using localization techniques.

The plan of the paper is as follows. In section \ref{sec:two} we introduce the class of accelerating black holes of interest
as well as present the supersymmetric and extremal limits. We also summarize the conditions required in order to get a regular
solution after uplifting to $D=11$. In section \ref{sec:Thermo} we discuss the thermodynamics of the accelerating black holes. Our analysis, which generalizes that of \cite{Anabalon:2018qfv} to include magnetic charge, actually covers the general class of accelerating black holes, without imposing the conditions required for regularity. In particular, 
by varying the parameters appearing in the solution, we derive a first law which involves introducing the tensions of the cosmic strings as extensive variables, along with their associated chemical potentials. In addition we also show that the first law can be formally extended to allow for variations of the cosmological constant, again generalising \cite{Anabalon:2018qfv}. In section \ref{sec:SUSYBHs} we introduce and study the complex locus of supersymmetric black hole solutions 
that includes the real supersymmetric and extremal black hole solution as a special case. This section also derives the
entropy function on this complex locus. In section \ref{sec:boundary} we obtain the boundary Killing spinors by directly solving the conformal Killing spinor equation. We construct the boundary supersymmetric Killing vector as a spinor bilinear and make contact with the results of \cite{BenettiGenolini:2019jdz}. We conclude with some discussion in section \ref{sec:disc}. Finally, in appendix \ref{appa} we show that
for supersymmetric solutions, the boundary metric and gauge field can be recast in a canonical form as studied in \cite{Hristov:2013spa}.

\section{The black hole solutions}\label{sec:two}

\subsection{The solutions}\label{sec:PD}

We consider solutions to minimal $D=4$, $\mathcal{N}=2$ gauged supergravity, which has a bulk action
given by
\begin{align}\label{act}
S_\mathrm{bulk}\, =\, \frac{1}{16\pi G_{(4)}}\int \dd^4x\sqrt{-g}\Big(R+\frac{6}{\ell^2}-F^2\Big)\,.
\end{align}
Here $F=\dd A$, and the cosmological constant is $-3/\ell^2<0$. 
Our starting point is the black hole solution to the corresponding equations of motion, given by 
\cite{Plebanski:1976gy,Podolsky:2006px} and discussed in
\cite{Ferrero:2020twa}
  \begin{align} \label{PDmetric}
\dd s^2 & =\, \frac{1}{H^2}\bigg\{
 -\frac{Q}{\Sigma}\Big(\frac{1}{\bk}\dd t- a\sin^2\theta\,\dd\phi \Big)^2 +\frac{\Sigma}{Q}\,\dd r^2   \nn
& \qquad \qquad  \qquad \qquad \qquad \qquad \quad 
+\frac{\Sigma}{P}\dd\theta^2 +\frac{P}{\Sigma}\sin^2\theta \Big( \frac{a}{\bk} \dd t -(r^2+a^2)\dd\phi \Big)^2  
 \bigg\}\, , 
\end{align}
where 
\begin{align}\label{PDfunctions}
 P(\theta)& \, = \, 1-2\alpha m\cos\theta +\Big(\alpha^2(a^2+e^2+g^2)- \frac{a^2}{\ell^2}\Big)\cos^2\theta \,,\nn
Q(r) &\, = \, (r^2-2m r + a^2+e^2+g^2)(1-\alpha^2 r^2) + \frac{r^2}{\ell^2}(a^2+r^2)\, ,\nn
H(r,\theta)&\, = \, 1-\alpha\, r\cos\theta\,, \nn
\Sigma(r,\theta)&\, = \, r^2+a^2\cos^2\theta \,,
\end{align}
and the gauge field is given by
\begin{align} \label{PDgauge}
A & \, =\, -e\frac{r}{\Sigma}\Big(\frac{1}{\bk}\dd t-a\sin^2\theta \dd\phi\Big)+g\frac{\cos\theta}{\Sigma}\Big(\frac{a}{\bk} \dd t-(r^2+a^2)\dd\phi\Big)\nn
& \,  = \, A_t\, \dd t + A_\phi \, \dd\phi\, . 
\end{align}
The solution depends on five free parameters $m,e,g,a$ and $\alpha$, loosely corresponding to mass, 
electric charge, magnetic charge, angular momentum and acceleration, respectively,
together with the $AdS$ radius $\ell>0$ and the 
constant $\bk>0$. The latter, which at this stage is a trivial constant that can be absorbed in a rescaling of the time coordinate, will be fixed later so that the Killing vector $\partial_t$ is appropriately normalized
in order to obtain a first law of thermodynamics. 
 We will focus on the case $m>0$. By utilising discrete isometries,
as discussed in \cite{Ferrero:2020twa}, without loss of generality in the physical Lorentzian solution we can consider 
 \begin{align} \label{signchoice}
 m \, > \,  0\, , \quad \mbox{and} \quad \alpha,e,g, a\, \ge \, 0\,,
  \end{align}
and for the most part we will take
\begin{align}
\alpha\, > \, 0\, .
\end{align}
The range of the $\theta$ coordinate is taken to be $0\le\theta\le\pi$. 
The black hole horizon is located at $r=r_+$, with $Q(r_+)=0$, where $r_+$ is the largest root of $Q$, and we demand 
\begin{align}\label{rprange}
0\, < \, r_+ \, < \, 1/\alpha\,.
\end{align}
The first inequality ensures that we avoid the black hole singularity at $r=0$ and the second that the horizon does not touch
the conformal boundary, as we will see later.
For convenience, we now continue with
\begin{align}
\ell\, =\, 1\,,
\end{align}
although we will briefly reinstate the $AdS$ radius $\ell$ when we discuss the first law of thermodynamics in section \ref{fstlaw}. 
This may be done via dimensional analysis, noting that the parameters $m,e,g,a$ and $1/\alpha$ all 
have dimensions of length.

The metric has two Killing vectors, $\partial_t$, $\partial_\phi$. Furthermore, the gauge we are using is such that
$\mathcal{L}_{\partial_t}A=\mathcal{L}_{\partial_\phi}A=0$. As in \cite{Ferrero:2020twa} we will discuss residual gauge transformations of the form 
\begin{align}\label{gaugets}
A \, \to \, \tilde A\, =\, A+\alpha_1 \dd t+\alpha_2 \dd\phi\,.
\end{align}
There is no choice of $\alpha_1,\alpha_2$ which makes $A$ globally well-defined when the magnetic charge is non-vanishing, $g\ne0$. 
As we will discuss later there is a natural choice of $\alpha_1$ which, after setting $g=0$, does make $A$ globally defined in the Euclidean 
solution.

\subsection{Regularity}\label{regresults}
When  $\alpha m \neq 0$, it is not possible to choose the period, $\Delta\phi$, of $\phi$ so that we obtain 
a smooth metric on a two-sphere $S^2$ on the surfaces of constant $t$ and  $r$. Instead, there is always a conical deficit 
at one or both of the poles $\theta=\theta_\pm$, where it is convenient to define
\begin{align}
\theta_- \, = \, 0\, , \qquad \theta_+ \, = \, \pi\, .
\end{align}
To see this, we introduce
\begin{equation}
\begin{split}
P_\pm  & \, \equiv \, P(\theta_\pm) \,  =\,  \Xi \pm 2\alpha m \, ,
\end{split}
\end{equation}
where 
\begin{align}\label{Xidef}
\Xi & \, \equiv \, 1 + \alpha^2(a^2+e^2+g^2) - {a^2}\, .
\end{align}
We then take the period $\Delta\phi$ of $\phi$ to be \cite{Ferrero:2020twa}
\begin{align}
\frac{\Delta\phi}{2\pi} \, =\,  \frac{1}{n_+ P_+}\, =\, \frac{1}{n_- P_-}\,.
 \end{align}
The constant $t$, $r$ surfaces, which we 
call $\SSigma$, are then topologically two-spheres but with conical deficit angles 
$2\pi(1-\frac{1}{n_\pm})$ at the poles $\theta=\theta_\pm$. If $n_\pm$ 
are coprime positive integers, then $\SSigma=\mathbb{WCP}^1_{[n_-,n_+]}$ 
is a weighted projective space, which is an orbifold also commonly known as a spindle. 

It will also be convenient to define the tensions of the associated ``cosmic strings"
\begin{align}\label{mupmdef}
\mu_\pm & \,  \equiv \, \frac{1}{4G_{(4)}}\left[1 -(\Xi \pm 2 \alpha m)\frac{\Delta\phi}{2\pi}\right]\, ,
\end{align}
so that 
\begin{align}\label{mupm}
\mu_- - \mu_+ & \, =\,  \frac{\alpha m}{G_{(4)}}\frac{\Delta\phi}{2\pi}\, ,\nn
\mu_-+\mu_+ & \, = \, \frac{1}{2G_{(4)}}\left(1- \Xi \frac{\Delta\phi}{2\pi}\right)\, ,
\end{align}
and
\begin{align}\label{npmmupm}
\frac{1}{n_\pm} & \,=\,  1  -4G_{(4)}\mu_\pm  \, .
\end{align}
The first equation in \eqref{mupm} shows immediately that 
we cannot take $n_-=n_+=1$, to obtain $\SSigma=S^2$, when $\alpha m\neq 0$. 
The orbifold Euler characteristic of $\SSigma$ is 
\begin{align}\label{euler}
\chi \, = \, \chi(\SSigma) \, = \,  \frac{1}{n_-}+\frac{1}{n_+} \, = \,  2-4G_{(4)}(\mu_-+\mu_+)\, .
\end{align}
Regarding $\mu_\pm$ (or equivalently $n_\pm$) as fixed, one can regard the first 
equation in \eqref{mupm} as fixing $\Delta\phi$ in terms of the parameters, while 
the second equation in \eqref{mupm}  is then a constraint on the parameters. 

Any solution of minimal $D=4$, $\mathcal{N}=2$ gauged supergravity automatically 
uplifts \emph{locally} to a solution of $D=11$ supergravity on an arbitrary Sasaki-Einstein seven-manifold $Y_7$
\cite{Gauntlett:2007ma}. In \cite{Ferrero:2020twa} it was shown that, 
starting from one of the black hole solutions with $\SSigma=\mathbb{WCP}^1_{[n_-,n_+]}$ 
a spindle with coprime positive integers $n_->n_+$, one can uplift to a completely smooth\footnote{Apart from the black hole singularity.}
solution of $D=11$ supergravity, free from conical deficit singularities, 
provided two conditions hold. Firstly, we require 
\be\label{g_is_malpha}
g \, = \, \alpha m\, .
\ee
Strictly speaking this is a sufficient condition, rather than necessary, but we shall see in the
next subsection that it is also required for supersymmetry. Secondly, 
the internal Sasaki-Einstein manifold $Y_7$ needs to be in the so-called regular class, {\it i.e.} 
$Y_7$ is the total space of a principal circle bundle over a positively curved K\"ahler-Einstein 
six-manifold. The precise circle fibration is in turn determined by the integers $n_\pm$ and the 
Fano index of the K\"ahler-Einstein base, 
as discussed in detail in \cite{Ferrero:2020twa}. We will not need any of 
the details of this uplift in the remainder of the paper, and will work entirely 
in four dimensions.

\subsection{Supersymmetry and extremality}\label{susyresults}

In this section we summarize the additional conditions on the parameters required for 
the solution to be supersymmetric, and also for the black hole solution to be 
extremal {\it i.e.} to have zero surface gravity. We follow \cite{Ferrero:2020twa} and \cite{Klemm:2013eca}. 

A solution to the equations of motion resulting from the action \eqref{act}
 is supersymmetric if there is a 
Dirac spinor $\epsilon$ satisfying the Killing spinor equation
\begin{align}\label{KSE}
\nabla_\mu\epsilon \, =\,  \left({\ii} A_\mu - \frac{1}{2}\Gamma_\mu - \frac{\ii}{4}F_{\nu\sigma}\Gamma^{\nu\sigma}\Gamma_\mu\right)\epsilon\, .
\end{align}
Here $\nabla_\mu$ is the spin connection and $\{\Gamma_\mu,\Gamma_\nu\} = 2g_{\mu\nu} $. 
Substituting the solutions of section \ref{sec:PD} into the 
integrability condition for the Killing spinor equation \eqref{KSE} leads to the following 
constraints when $\alpha\neq 0$:
\begin{align}\label{SUSY}
g & \, = \, \alpha m\, , \nn
0 & \, = \, \alpha^2(e^2+g^2)(\Xi + a^2)-(g- a \alpha e)^2 \, .
\end{align}
We refer to these as the supersymmetry equations, the first of which was discussed in 
a quite different context at the end of the previous subsection. 

A supersymmetric solution is also extremal provided the following relation also holds \cite{Ferrero:2020twa}
\begin{align}\label{extremal}
a g^2 (a \alpha e  -g)(e+a\alpha g)+ \alpha^3 e^2 (e^2 +g^2)^2 &=\, 0 \, .
\end{align}
The first equation in \eqref{SUSY} is of course straightforward to implement, 
although imposing the second equation together with \eqref{extremal} is 
at best cumbersome. We shall see later in section \ref{sec:SUSYBHs} 
that these equations, and indeed also various physical quantities of interest, significantly simplify 
if one first introduces a different set of variables.

\section{Thermodynamics}\label{sec:Thermo}

In this section we use holography to determine the thermodynamics of the black holes. 
Our results generalize those given in \cite{Anabalon:2018qfv} to also include non-vanishing
magnetic charge, $g\ne 0$, which is essential for supersymmetry.

\subsection{Boundary stress tensor, current and conserved charges}\label{sec:onshact_charges}

As discussed in \cite{Anabalon:2018qfv,Ferrero:2020twa}, the conformal boundary of the metric \eqref{PDmetric} is 
located at $H(r,\theta)=0$. It is then  convenient to introduce a new radial coordinate $z$ via\footnote{It would be less convenient to write $r=\frac{1}{\alpha\cos\theta}+z$, in particular at $\theta=\pi/2$.}
\be
\frac{1}{r} \, = \,  \alpha \cos \theta + z\,,
\ee
so that the conformal boundary is located at $z=0$. We thus introduce a small cutoff $\epsilon \geq0$ and study the near-boundary hypersurfaces of constant $z$ in the limit $z = \epsilon \to 0$. 
Notice then that 
the condition \eqref{rprange} ensures that the black hole horizon does not touch the conformal boundary at $\theta=0$.
In this parametrization, the four-dimensional coordinates are $x^\mu = (t,\theta,\phi,z)$, while the coordinates on the hypersurfaces are $x^i = (t,\theta,\phi)$. The ADM decomposition of the bulk metric is
\be
\dd s^2 \, = \, N^2 \dd z^2 + h_{ij}(\dd x^i + N^i \,\dd z)(\dd x^j + N^j \,\dd z)\,,
\ee
where the induced metric $h_{ij}$, the lapse function $N$ and the shift vector $N^i = (0,N^\theta,0)$ depend both on $z$ and $\theta$. The outward-pointing unit vector normal to the hypersurfaces of constant $z$ is given by
\be
n \, = \,   \frac{1}{N} \left( N^i \partial_i -\partial_z\right)\,.
\ee
The extrinsic curvature of the hypersurfaces, as a tensor on the hypersurface, is given by 
\be\label{extcurv}
K_{ij} \,= \, \frac{1}{2} \mathcal{L}_n g_{ij}\, = \, - \frac{1}{2N} \left( \partial_z h_{ij} - \nabla_i^{(h)} N_j - \nabla_j^{(h)}N_i\right)\,,
\ee
where here $\nabla_i^{(h)}$ is the Levi-Civita connection of $h_{ij}$. 

The metric on the conformal boundary is defined to be 
\begin{align}\label{metric_bdry}
\dd s^2_{\rm bdy} \, &\equiv \, \lim_{\epsilon\to 0} \, \epsilon^2 h_{ij}|_{z=\epsilon} \, \dd x^i \dd x^j\nn
=&
-\tilde P\Big(\frac{1}{\bk}\dd t -\frac{a(1-\alpha^2 P)\sin^2\theta}{\tilde P}   \dd\phi\Big)^2 
+
\frac{\big(1+a^2 \alpha ^2 \cos ^4\theta \big)^2}{P \big(\tilde P+a^2 \alpha ^2 \cos ^4\theta\big)}
\dd\theta^2\nn
&+
{P \sin ^2\theta \Big(1+\frac{a^2 \alpha ^2 \cos ^4\theta}{\tilde P}\Big)}
\dd\phi^2\,,
\end{align}
and we have defined
\begin{align}\label{tildePdef}
\tilde P \, = \, \tilde{P}(\theta)\, \equiv \, 1-\alpha^2 P(\theta)\sin^2\theta\,.
\end{align}
In the $a=0$ limit we recover the boundary metric studied in some detail in section 6 of \cite{Ferrero:2020twa}, 
up to a conformal factor. 
We emphasize that the Cotton tensor of the boundary metric \eqref{metric_bdry} is, generically,
non-vanishing, and hence the boundary is not conformally flat. 
This is in contrast to the case when the acceleration parameter vanishes, $\alpha = 0$, when the Cotton tensor vanishes
 and
the boundary is conformally flat.\footnote{The Cotton tensor also vanishes when $m=g=e=0$.}
The boundary gauge field is defined to be
\begin{align}\label{gfdexpbdy}
A_{\rm bdy} \, &= \, \lim_{\epsilon\to 0}\, A_i|_{z=\epsilon}\, \dd x^i \\
\, &= \, -\frac{\cos\theta}{1+\alpha^2 a^2\cos^4\theta} \left[ \frac{\alpha}{\bk}\left(e- g \alpha a \cos^2\theta\right) \dd t + \left(g+g \alpha^2a^2\cos^2\theta - e\alpha a \sin^2\theta \right)\dd\phi  \right]\, .\nonumber
\end{align}
Note that in the gauge we are using, while $A^{\rm bdy}_i=A^{\rm bdy}_i(\theta)$, we have $\int_0^\pi \dd\theta A^{\rm bdy}_i=0$.

To calculate the stress tensor of the boundary theory we need to consider the total action given by
\begin{align}
S\, =\, S_\mathrm{bulk}+S_\mathrm{bdy}\,.
\end{align}
Here the bulk action is given in \eqref{act}, while the boundary action, which includes the Gibbons-Hawking term as
well as the counterterms, is given by
\begin{align}
S_\mathrm{bdy}\, =\, \frac{1}{16\pi G_{(4)}}\int_{\mathrm{bdy}} \dd^3x\sqrt{-h}\,\big(2K - 4 -R(h)\big)\,.
\end{align}
Here $K = h^{ij} K_{ij}$ is the trace of the extrinsic curvature, and $R(h)$ is the Ricci scalar of the metric $h_{ij}$.

The renormalized energy-momentum tensor is given by \cite{Balasubramanian:1999re}
\be
T_{ij} \,=\, \frac{1}{8\pi G_{(4)}}\,\lim_{\epsilon\to 0}\, \frac{1}{\epsilon}\Big[ -K_{ij} + h_{ij} K -2 h_{ij} + R_{ij}(h)- \frac{1}{2}h_{ij} R(h) \Big]_{z=\epsilon}\,.
\ee
The explicit expression is lengthy and so we will not report it here.
Similarly the electric current of the dual field theory is defined by
\begin{align}
j^i&\, =\, -\frac{1}{4\pi G_{(4)}}  \lim_{\epsilon\to 0} \left[\frac{1}{\epsilon^3} n_\mu F^{\mu i}\right]_{z=\epsilon}\,,
\end{align}
and explicitly we have
\begin{align}
j^t &\, =\,  \bk\frac{e\left[ 1+3\alpha^2a^2 x^2 - \alpha^2a^2x^4(3+\alpha^2a^2x^2)  \right]  + g\alpha a \left[1-3x^2 -\alpha^2a^2x^4(3-x^2)  \right]}{4\pi G_{(4)}(1+\alpha^2a^2x^4)^3}\,,\nn[2mm]
j^\phi &\, =\,  \frac{\alpha[e\,\alpha a x^2 \left( 3-\alpha^2a^2x^4 \right)  + g \left(1-3\alpha^2a^2x^4 \right)]}{4\pi G_{(4)}(1+\alpha^2a^2x^4)^3}\,,
\end{align}
where we are using the variable
\begin{align}\label{xvar}
x \, \equiv \, \cos\theta\, ,
\end{align}
and we have $j^\theta=0$. 

One can check directly  that $j^i$ and $T_{ij}$ satisfy the Ward identities
\begin{align}
D_i j^i &\, = \, 0\,,\nn
D^i T_{ij} &\, = \, - j^i F^{\rm bdy}_{ij} \,,
\end{align}
where $D_i$ is the Levi-Civita connection of the boundary metric \eqref{metric_bdry} ({\it i.e.} the limit  $\epsilon\rightarrow 0$ of the metric $h_{ij}$ rescaled by $\epsilon^2$), which is also used to raise and lower indices,
and $F_{\rm bdy} = \dd A_{\rm bdy}$. Furthermore, we have checked that $T^i{}_i=0$, as expected. 
It is useful to also recall that if the boundary metric has a Killing vector field $k$, satisfying
$\mathcal{L}_kh_{ij}=0$ and  $\mathcal{L}_kA_\mathrm{bdy}=0$, then we obtain a conserved boundary current:
\begin{align}\label{conskvcond}
D_i[(T^i{}_j+j^i A^\mathrm{bdy}_j)k^j]\, =\, 0\,.
\end{align}
Note that the current $(T^i{}_j+j^i A^\mathrm{bdy}_j)k^j$ changes under gauge transformations that maintain $\mathcal{L}_kA_\mathrm{bdy}=0$, a point we return to below.

We are now in a position to compute the total mass and angular momentum in the boundary theory. These are conserved charges associated with Killing vectors of the boundary metric, and given by integrals at the boundary over a two-dimensional spatial hypersurface $\SSigma_\infty$ of constant time. However, some care is required in
choosing these constant time hypersurfaces. To proceed, we first introduce new coordinates on the boundary defined by
\begin{align}
t\, =\, \bar t\, ,\qquad \phi\, =\, \bar\phi+\frac{\Delta\phi}{2\pi}\Omega_\infty \bar t\,,
\end{align}
with 
\begin{align}\label{barredvecs}
\partial_{\bar t}\, =\, \partial_t+\Omega_\infty\frac{\Delta\phi}{2\pi}\partial_\phi\, ,\qquad \partial_{\bar\phi}\, =\, \partial_\phi\,,
\end{align}
and $\Omega_\infty$ is a constant which will be chosen momentarily. 
We denote the surfaces of constant $\bar t$ on the conformal boundary by $\SSigma_\infty$. The associated ADM decomposition of the boundary metric \eqref{metric_bdry} can be written 
\be
\dd s^2_{\rm bdy} \, = \,  - \nu^2 \dd \bar t^2 + \gamma_{\hat\imath\hat\jmath} \left(\dd x^{\hat\imath} + \nu^{\hat\imath} \dd \bar t\right)\left(\dd x^{\hat\jmath} + \nu^{\hat\jmath} \dd \bar t\right)\,,
\ee
where $\gamma_{\hat\imath\hat\jmath}$, $\hat\imath,\hat\jmath =1,2$, is the induced metric on $\SSigma_\infty$, 
and the future-directed unit vector normal to the hypersurfaces of constant $\bar t$ is 
\be
u \, = \,  u^i\, \partial_i \, = \, \frac{1}{\nu} \left(\partial_{\bar t} - \nu^{\hat\imath}\partial_{\hat\imath}\right)\,.
\ee
For a Killing vector $k$, with associated conserved current as in \eqref{conskvcond}, we can then define the conserved charge $Q_k$ as
\begin{align}
Q_k \, &\equiv \, \int_{\SSigma_\infty}  \dd^2x \,\sqrt{\gamma}\, u_{i} (T^i{}_j+j^i A^\mathrm{bdy}_j)k^j \,.
\end{align}

We now fix the choice of time coordinate $\bar t$ by defining
\begin{align}\label{ominftydef}
\Omega_\infty &= -\frac{2\pi}{\bk\Delta\phi}\frac{a(1-\alpha^2\Xi)}{\Xi (1+a^2\alpha^2)}\,.
\end{align}
The total energy or mass $M$ is associated with the boundary Killing vector $k=\partial_{\bar t}$. We define
\begin{align}
M&\, \equiv \, Q_{\partial_{\bar t}}\, \equiv \, \int_{\SSigma_\infty}  \dd^2x \,\sqrt{\gamma}\, u_{i} \left( T^i{}_{\bar t} + j^i A^\mathrm{bdy}_{\bar t} \right)
\, = \, \int_{\SSigma_\infty}  \dd^2x \,\sqrt{\gamma}\, u_{i}\,  T^i{}_{\bar t}\nn
&\, = \, \frac{m \Delta \phi}{2\pi \bk G_{(4)}}\frac{(\Xi+a^2)(1-\alpha^2\Xi)}{\Xi(1+\alpha^2a^2)}\,,
\end{align}
where the last equality in the first line is a feature of the gauge we are using.\footnote{This is not true, for example, for the special case of $g=0$
if we used a gauge transformation of the form \eqref{gaugets} with $\alpha_1$ chosen so that $\tilde A$ was globally defined; in fact one would find $Q_{\partial_{\bar t}}=M-\Phi_e Q_e$, with $\Phi_e$ and $ Q_e$ the chemical potential and electric charge appearing in the first law.}
Our definition of $M$ depends on the choice  $\Omega_\infty$; we will later see
that this definition of $M$ appears in the first law, with $M$ a function of the entropy $\SBH$ and the charges $J,Q_e$, defined below.
 In the special case that $\alpha=0$, it is the 
definition of mass that has appeared in previous discussions of the Kerr-Newman $AdS_4$ black holes ({\it e.g.} see \cite{Papadimitriou:2005ii}).
When $\alpha\ne 0$ and $g=0$, this choice was also used in \cite{Anabalon:2018ydc}. Finally, we will later see that
this definition of $M$ leads to a simple form for the BPS relation between $M$ and $J$, $Q_e$ for supersymmetric solutions.

The total angular momentum $J$ is associated with the boundary Killing vector
$ k=-\frac{\Delta\phi}{2\pi} \,\partial_\phi$, and we define it as
\begin{align}\label{Jdefn}
J \,&\equiv \, - \frac{\Delta\phi}{2\pi}\int_{\SSigma_\infty}  \dd^2x \,\sqrt{\gamma}\, u_{i} \left( T^i{}_{\phi} + j^i A^\mathrm{bdy}_\phi \right)\nn
&= \, - \frac{\Delta\phi}{2\pi}\int_{\SSigma_\infty}  \dd^2x \,\sqrt{\gamma}\, u_{i} \left( T^i{}_{\phi}  \right)\, 
=\, \frac{a m}{G_{(4)}} \left(\frac{\Delta\phi}{2\pi}\right)^2\,.
\end{align}
Note that $J$ is independent of $\Omega_\infty$ since $\partial_\phi=\partial_{\bar\phi}$.
In \cite{Ferrero:2020twa}, following \cite{Papadimitriou:2005ii}, it was emphasized that $J$ is a kind of Page charge, and in particular
using Stokes' theorem, it was also explained how $J$ can be obtained as an integral over the  horizon, which is a copy $\SSigma_H$ of 
the two-dimensional surface $\SSigma$. 

The definition of the conserved electric and magnetic charges is more straightforward, and in particular are gauge invariant.
The total electric charge, $Q_e$, is obtained by integrating the charge density over $\SSigma_\infty$, and we have
\begin{align}
Q_e \,&=\, -\int_{\SSigma_\infty}  \dd^2x \,\sqrt{\gamma}\, u_{i} j^i \,=\,\frac{1}{4\pi G_{(4)}}\int_{\SSigma_\infty} *_4F\,=\,  \frac{e\Delta\phi}{2\pi G_{(4)}}\,,
\end{align}
while the total magnetic charge, $Q_m$, is given by
\begin{align}\label{magflux}
Q_m\, \equiv\, \frac{1}{4\pi G_{(4)}}\int_{\SSigma_\infty} F_{\rm bdy}\, = \, \frac{g \Delta\phi}{2\pi G_{(4)}} \,.
\end{align}
Both $Q_e$ and $Q_m$ are independent of $\Omega_\infty$. Furthermore, like $J$, they can also be obtained as horizon integrals over 
$\SSigma_H$,
as discussed in \cite{Ferrero:2020twa}. 

\subsection{Black hole entropy and on-shell action}\label{sec:BHentropy}
The black hole horizon is located at $r=r_+$. The null generator of the black hole horizon, $\nullH$, 
is given by 
\be 
\nullH \, = \,  \partial_t + \Omega_H \frac{\Delta \phi}{2\pi} \,\partial_\phi\, ,
\ee 
where the angular velocity of the horizon, $\Omega_H$, is
\be\label{angvelh}
\Omega_H \, =  \, \frac{2\pi}{\bk\Delta\phi}\,\frac{a}{r_+^2 + a^2} \,.
\ee
If we introduce new coordinates 
\begin{align}\label{primedcoordinates}
t\, =\, t',\qquad  \phi\, =\, \phi'+\Omega_H \frac{\Delta \phi}{2\pi} t'\, ,
\end{align}
then the generator is $\nullH=\partial_{t'}$ and furthermore they are the natural coordinates to
show that the Euclidean metric, discussed below, is regular at the horizon.

The Bekenstein-Hawking entropy of the black hole is given by
\begin{align}
\SBH\, =\, \frac{\Delta\phi}{2G_{(4)}}\frac{r_+^2+a^2}{1-\alpha^2 r_+^2}\,.
\end{align}
The surface gravity, $\kappa_{sg}$, is obtained via $\kappa_{sg}^2 =  -\frac{1}{2}\nabla_\mu \nullH_\nu \nabla^\mu \nullH^\nu$, evaluated at the horizon. Identifying the temperature via $T= \beta^{-1}= \frac{\kappa_{sg}}{2\pi}$,  
we have 
\begin{align}
 T\, = \, \frac{Q'(r_+)}{4\pi\bk (a^2+r_+^2)}\,.
\end{align}

To evaluate the on-shell action we perform the Wick rotation 
\begin{align}\label{wick}
t\, =\, -\ii\tau\,,\qquad  I\, =\, -\ii S\,.
\end{align}
To get a real solution we should also take $a=\ii a_E$ and $e=\ii e_E$.
Moving to new coordinates $\tau=\tau'$ and $\phi=\phi'-\ii\Omega_H \frac{\Delta \phi}{2\pi}\tau' $ so that
$\nullH=\ii\partial_{\tau'}$, we find that the metric is smooth at $r=r_+$ provided that 
we identify 
\begin{align}
(\tau',\phi')\, \sim  \, (\tau'+\beta m_1,\phi'+\Delta\phi \, m_2)\, ,
\end{align}
where $m_i\in\mathbb{Z}$. Equivalently, in the unprimed 
Euclidean coordinates we have the twisted identification
\begin{align}\label{twistedidentsrev}
(\tau,\phi)\, \sim \, (\tau+\beta m_1, \phi-\ii\Omega \frac{\Delta \phi}{2\pi}\beta m_1+\Delta\phi\,  m_2)\,,\qquad  m_i\in\mathbb{Z}\,.
\end{align}
While $\tau$ is not a periodic coordinate, it is useful to note that we have $\int \dd\tau \dd\phi=\beta\Delta \phi$.
We note that the Euclidean solution has topology $\R^2\times \SSigma$, where 
the horizon $r=r_+$ is at the origin of $\R^2$, and the Euclidean time $\tau'$ plays the role of a polar coordinate on $\R^2$.
Indeed, it will sometimes be convenient to use the canonically normalized coordinates defined by
\begin{align}
\psi \, \equiv \, \frac{2\pi}{\beta}\tau'\, , \qquad \ssigma \, \equiv \, \frac{2\pi}{\Delta\phi}\phi'\, ,
\end{align}
so that $\psi$ is $(2\pi)$-period polar coordinate on the $\R^2$ normal to the horizon, and $\ssigma$ is a $(2\pi)$-period 
coordinate on the surface $\SSigma$. We would also like to highlight that the gauge field is not regular at the black hole horizon; 
indeed, in general, there is no gauge in which this is the case. In the special case that $g=0$, it is possible if we make a gauge transformation of the form \eqref{gaugets} with $\alpha_1=(er_+)/[\bk(a^2+r_+^2)]$.

After some calculation we find that the Euclidean on-shell action (see equation \eqref{wick}) 
is
\begin{align}\label{ios}
I & \,  = \,   
\frac{\beta \Delta \phi}{16\pi \bk G_{(4)}}  \Bigg[ -  4 r_+\! \left(\frac{a^2+r_+^2}{(\alpha^2r_+^2-1)^2} + \frac{e^2-g^2}{a^2+r_+^2}\right)\nn
& \qquad \qquad \qquad \qquad \qquad \qquad \qquad \qquad + 4m \left(1-2\alpha^2 -\frac{2\alpha^4(e^2+g^2)}{1+\alpha^2a^2} \right)\Bigg]\,. 
\end{align}
We next define the electrostatic potential, $\Phi_e$ via
\begin{align}
\Phi_e\, \equiv \, \Phi_\infty -\Phi_H\,,
\end{align}
with $\Phi_H=\left.V\cdot A\right|_{r\to r_+}$, which is necessarily a constant, and $\Phi_\infty$ to be the $\theta$-independent component
of $V\cdot A_{\text{bdy}}$ ({\it i.e.} the zero mode). In the gauge we are using, as noted below \eqref{gfdexpbdy}, 
we have $\Phi_\infty=0$ and so
\be\label{phie}
\Phi_e \, = \, \frac{e\,r_+ }{\bk(r_+^2 + a^2)} \,.
\ee
We then we immediately find that the following quantum statistical relation is satisfied:
\be\label{qsrresult}
\begin{split}
I & 
\, = \,   -\SBH + \beta(M- \Omega J - \Phi_e Q_e)\, ,
\end{split}
\ee
where we have defined
\begin{align}
\Omega \, \equiv \,  \Omega_H - \Omega_\infty\,.
\end{align}

A number of comments are now in order. First, a derivation of the expression for $\Phi_e$ in \eqref{phie} was given in 
appendix E of \cite{Ferrero:2020twa}, following \cite{Anabalon:2018qfv}; the factor of $\bk$ here arises because of the normalization
of the time coordinate in \eqref{PDmetric}. 
Second, to discuss the gauge transformations \eqref{gaugets} it is illuminating to rewrite \eqref{qsrresult} in the form
$I=-\SBH + \beta(M+ \Omega_\infty J-\Phi_\infty Q_e)-\Omega_H J + \Phi_H Q_e\,$, where as we noted above in the current gauge 
$\Phi_\infty=0$. In particular we notice that $M+ \Omega_\infty J-\Phi_\infty Q_e $ is the conserved charge associated with the 
Killing vector $\partial_t$, {\it i.e.} $Q_{\partial_t}$.
We first consider the gauge transformations as in \eqref{gaugets} that are parametrized by $\alpha_2$: as shown in \cite{Ferrero:2020twa}
this takes $J\to J+\alpha_2 Q_e$ and $\Phi_H\to \Phi_H+\alpha_2\Omega_H$, with $\Phi_\infty$ invariant (and, of course, $Q_e$ also). Furthermore, it is clear that $M+ \Omega_\infty J-\Phi_\infty Q_e$ is invariant because it is the conserved charge associated with the 
Killing vector $\partial_t$ and the $A_t$ component is unchanged for this gauge transformation. 
Thus, we see that $I$ is invariant, as it had to be.
For the gauge transformations parametrized by $\alpha_1$, we have 
$M+ \Omega_\infty J-\Phi_\infty Q_e$ will transform to the same thing plus $-\alpha_1 Q_e$ while
$\Phi_H\to\Phi_H+\alpha_1$, again leaving $I$ invariant.
Importantly, as we show in the next section, within the gauge we are using, there is a standard first law for
$M=M(\SBH,J,Q_e)$, for a suitably chosen $\kappa$, and, with the quantum statistical relation we can then deduce $I=I(T,\Omega,\Phi_e)$ and hence
$W=T I$ can be identified as the Gibbs free energy of the dual field theory. Finally, 
we emphasize that we have not yet used the regularity constraints of section \ref{regresults}, nor those for supersymmetry as
in section \ref{susyresults}. 

\subsection{The first law}\label{fstlaw}

Elsewhere we have set the $AdS$ radius $\ell=1$. We now briefly restore $\ell$, so it can temporarily be varied. 
We first collect formulas that we have derived so far:
\begin{align}\label{allquantities}
M & \, = \, \frac{m \Delta \phi}{2\pi \bk G_{(4)}}\frac{(\Xi+a^2/\ell^2)(1-\alpha^2\ell^2\Xi)}{\Xi(1+\alpha^2a^2)}\, , \quad 
\SBH  \, = \, \frac{\Delta\phi}{2 G_{(4)}}\frac{r_+^2+a^2}{1-\alpha^2 r_+^2}\, ,\nn
Q_e & \, = \, \frac{e \Delta\phi}{2\pi G_{(4)}}\, ,\qquad \qquad \qquad \qquad \qquad  \qquad  Q_m \,  = \, \frac{g \Delta\phi}{2\pi G_{(4)}}\, ,\nn
\Phi_e & \, = \,  \frac{e r_+}{\bk(r_+^2+a^2)}\, , \qquad \qquad \qquad \qquad \quad \quad \ \Phi_m  \, = \,  \frac{g r_+}{\bk(r_+^2+a^2)}\, , 
\end{align}
as well as 
\begin{align}
J & \, = \, \frac{am}{G_{(4)}}\left(\frac{\Delta\phi}{2\pi}\right)^2\, , \qquad \qquad\quad  \qquad \qquad \ \Omega_H \, = \, \frac{2\pi}{\bk \Delta\phi}\frac{a}{r_+^2+a^2}\, ,\nn
\Omega_\infty &\, = \,  -\frac{2\pi}{\bk\Delta\phi}\frac{a(1-\alpha^2\ell^2\Xi)}{\Xi \ell^2(1+a^2\alpha^2)}\, , \qquad \qquad \qquad\ \   \Omega \,  = \,  \Omega_H - \Omega_\infty\, ,\nn
\mu_\pm & \, = \, \frac{1}{4G_{(4)}}\left[1 -(\Xi \pm 2\alpha m)\frac{\Delta\phi}{2\pi}\right]\, , \qquad \quad \ \ \, T \, = \,  \frac{Q'(r_+)}{4\pi \bk (r_+^2+a^2)}\, , 
\end{align}
where $\Xi  = 1 - \frac{a^2}{\ell^2} + \alpha^2(a^2+e^2+g^2)$.
Here we have also added the magnetic potential $\Phi_m$, which may in principle 
be derived in a similar manner to the electrostatic potential $\Phi_e$ discussed in the previous subsection. 
Alternatively, electric-magnetic duality simply exchanges the parameters $e$ and $g$, which 
leads to the form for $\Phi_m$ given in \eqref{allquantities}.  
It is interesting to note that 
\begin{align}
\Phi_m Q_e - \Phi_e Q_m =\, \,   0\,.
\end{align}

To state the  most general form of the
first law we also introduce the variables $\lambda_\pm$ conjugate to the 
cosmic string tensions $\mu_\pm$, together with the cosmological 
constant parameter $p$, its conjugate variable $v$ and the function $\xi$: 
\begin{align}\label{morequantities}
\lambda_\pm & \, = \, \frac{r_+}{\bk(1\pm \alpha r_+)}-\frac{m[\Xi+a^2/\ell^2+a^2/\ell^2(1-\alpha^2\ell^2\Xi)]}{\bk\, \Xi^2(1+a^2\alpha^2)}\mp \frac{\alpha\ell^2(\Xi+a^2/\ell^2)}{\bk(1+a^2\alpha^2)}\, ,\nn
p & \, = \,  \frac{3}{8\pi \ell^2 G_{(4)}}\, ,\nn
 v & \, = \,  \frac{4\pi}{3\bk}\frac{\Delta\phi}{2\pi}\left[r_+\frac{(r_+^2+a^2)}{(1-\alpha^2r_+^2)^2} 
+ m\frac{a^2(1-\alpha^2\ell^2\Xi)+\alpha^2\ell^4\Xi(\Xi+a^2/\ell^2)}{\Xi(1+a^2\alpha^2)}\right]\, ,\nn
\xi \,&=\, M - T\SBH - \Phi_eQ_e- \Phi_mQ_m -\Omega J + \lambda_+\mu_+ + \lambda_-\mu_- - pv\,.
\end{align}
See {\it e.g.}~\cite{Caldarelli:1999xj,Kastor:2009wy,Kubiznak:2016qmn} for a discussion of $p,v$ in black hole thermodynamics with a cosmological constant.
In addition to the  quantum statistical relation \eqref{qsrresult}, we find 
that the following Smarr relation holds:
\begin{align}\label{Smarr_rel}
 M &\, = \,  2(T\SBH + \Omega J - p v)+ \Phi_e Q_e + \Phi_m Q_m\, ,
\end{align}
which generalizes the result of \cite{Anabalon:2018qfv} to include magnetic charge.
Similarly to \cite{Anabalon:2018qfv}, we now find that provided we choose the normalization of the 
time coordinate by setting
\be\label{kappa}
\bk \, = \,  
\frac{\sqrt{(\Xi+a^2/\ell^2)(1-\alpha^2\ell^2 \Xi)}}{1+a^2\alpha^2}\, ,
\ee
then the following first law holds:
\begin{align}
\dd M & \, = \, T \dd\SBH  + \Phi_e \dd Q_e + \Phi_m \dd Q_m + \Omega\, \dd J - \lambda_+ \dd\mu_+ - \lambda_- \dd \mu_- \nn
& \qquad + v\dd p - \xi \frac{\dd G_{(4)}}{ G_{(4)} }\, ,
\end{align}
where we vary with respect to all seven parameters $m,a,e,g,\alpha,\Delta\phi,\ell$ and we have also entertained
the possibility of also allowing for variations of the Newton's constant $G_{(4)}$ as in \cite{Visser:2021eqk,Cong:2021fnf}.

Turning now to a standard holographic perspective we want to keep $\ell,G_{(4)}$ fixed, and from now on we will again continue with
\begin{align}
\ell\, =\, 1\, .
\end{align}
In addition we would like to keep $\mu_\pm$ fixed, as these determine the conical deficit angles  
 and thus the topology of the surfaces $\SSigma$.
If we also keep the magnetic charge $Q_m$ fixed,  so as to fix all topological data at the boundary, 
then we recover a more standard first law 
\begin{align}\label{standardfirstlaw}
\dd M \, = \, T\dd \SBH + \Phi_e \dd Q_e + \Omega\, \dd J \, ,
\end{align}
where this now holds for a three-parameter family of solutions. In practice, recalling the explicit expressions of $Q_m,\mu_\pm$ in \eqref{allquantities}, we see that holding these quantities fixed means that we are fixing the following products of parameters: $g\Delta\phi$, $\alpha m \Delta\phi$, $\Xi\Delta\phi$. 

Using the version \eqref{standardfirstlaw} of the first law in the variation of the quantum statistical relation \eqref{qsrresult}, one obtains the variation of the on-shell action
\begin{align}\label{variation_action}
\dd I &\, = \,  \dd\beta (M - \Omega J - \Phi_e Q_e) +  \beta(  - \dd\Omega J  - \dd\Phi_e Q_e )\nn
&\, = \,  \dd\beta M - \dd(\beta\Omega) J - \dd(\beta\Phi_e)  Q_e \,,
\end{align}
expressing the fact that the on-shell action can be viewed as a function of the chemical potentials, $I=I(\beta, \Omega,\Phi_e)$, such that
\be
\left.\frac{\partial I}{\partial\beta}\right|_{\beta\Omega,\,\beta\Phi_e} \, = \,  M\,,\qquad -\frac{1}{\beta}\left.\frac{\partial I}{\partial\Omega}\right|_{\beta,\,\Phi_e} \, = \,  J\,,\qquad-\frac{1}{\beta}\left.\frac{\partial I}{\partial\Phi_e}\right|_{\beta,\,\Omega} \, = \, Q_e\,.
\ee
Equivalently, one can express the variation of the free energy $W = I/\beta$ as
\be\label{variation_free_energy}
\dd W \,=\, -\SBH\, \dd T -J \, \dd\Omega - Q_e\, \dd\Phi_e\, ,
\ee
and so $W=W(T,\Omega,\Phi_e)$.

We emphasize the versions of the first law given in
\eqref{standardfirstlaw}, \eqref{variation_action} or \eqref{variation_free_energy} all utilize the expression for $\kappa$ given in \eqref{standardfirstlaw}. 
Notice that when there is no acceleration, $\alpha=0$, we have $\kappa=1$ and hence the role of $\kappa$ in the first law
is intrinsically connected with the acceleration. It would be interesting to have a better understanding of $\kappa$ 
and the following observation may be useful:
if we vary the boundary metric components appearing\footnote{Note that the coordinates used in 
\eqref{metric_bdry} are subject to the identifications given in \eqref{twistedidentsrev}.}
 in \eqref{metric_bdry} with respect to the parameters
(with fixed spindle data) then we have $\int_\text{bdy}\diff^3x\sqrt{-h}T^{ij}\delta h_{ij}=0$.

Finally, we note that we can choose the fixed boundary data $\mu_\pm$, $Q_m$ so that the surface $\SSigma$ is a spindle, $\SSigma=\mathbb{WCP}^1_{[n_-,n_+]}$, and the $D=11$ solution is regular on and outside the horizon. To achieve this, we need to 
demand that $\mu_\pm$ are chosen so that $n_\pm$ given by \eqref{npmmupm} are integer. We also
need to impose the condition \eqref{g_is_malpha}, {\it i.e.} $g=\alpha m$; 
using the first equation in \eqref{mupm} this is equivalent to the following relation between our fixed quantities,
\be\label{balancing_Qm_Deltamu}
G_{(4)}Q_m \,=\, G_{(4)}(\mu_- - \mu_+ )\,=\, \frac{n_--n_+}{4 n_- n_+} \,.
\ee
This is a balancing condition between the magnetic charge $Q_m$ and the relative conical deficit angles between the poles of $\SSigma$. With this choice of boundary data, we obtain a black hole with spindle horizon depending on three continuous parameters and satisfying the thermodynamics discussed in this section, in particular the quantum statistical relation \eqref{qsrresult} and the first law \eqref{standardfirstlaw}. 
 
\section{Supersymmetric and extremal black holes}\label{sec:SUSYBHs}

We now turn to examine the supersymmetry \eqref{SUSY} and extremality 
\eqref{extremal} conditions in detail. Some of the thermodynamic quantities of interest were
 computed in \cite{Ferrero:2020twa} for supersymmetric and extremal solutions, 
although this involved first going to the near horizon limit, where the solution simplifies. 
This indirect method was 
used due to the unwieldy nature of imposing 
\eqref{SUSY} and \eqref{extremal}, which depend on the original set of parameters 
$m,e,g,a,\alpha$ (but not $\Delta\phi$).  We begin in this section by introducing a 
new set of parameters, in which both these equations and the physical quantities 
of interest take a much simpler form. 
As we shall see, imposing only supersymmetry leads naturally
to complex parameters, which then describe an analytic continuation of the solutions, and the parameters become real for extremal solutions.

For this supersymmetric and complex family of solutions we will derive an explicit expression for the
action, which is complex, and show that it can be expressed in terms of suitably defined complex chemical potentials.
Moreover, we show that after a Legendre transform we obtain an expression for the entropy in terms of
the conserved charges and furthermore that we obtain the correct expression for the entropy of the (real) extremal and supersymmetric black holes by demanding that the resulting expression is real.

\subsection{New variables}\label{sec:newvars}

We begin by defining
\begin{align}\label{mudef}
\mu \, \equiv \, \frac{1-2G_{(4)}(\mu_-+\mu_+)}{2G_{(4)}(\mu_--\mu_+)} \, = \, \frac{n_-+n_+}{n_--n_+}\, ,
\end{align}
in terms of the cosmic string tensions $\mu_\pm$ introduced in \eqref{mupmdef}, or equivalently 
spindle parameters $n_\pm$ in \eqref{npmmupm}. Eqs.~\eqref{mupm} together with the condition \eqref{g_is_malpha} then imply
 \begin{align}
\label{cond1}
\Xi \,  = \, 2 g\mu\,,\qquad g  \, = \, \alpha m\,, 
 \end{align}
where recall that the second equation is related to regularity 
of the uplifted solution in $D=11$, but is also the first 
supersymmetry condition in \eqref{SUSY}. 
 Using \eqref{npmmupm} we may also rewrite \eqref{balancing_Qm_Deltamu} as
\begin{align}\label{gqmexps}
G_{(4)}Q_m 
 \, = \,  \frac{n_--n_+}{4 n_+n_-}\, = \, \frac{1}{2 n_+(1+\mu)}\,.
 \end{align}
Note that expressions containing $(n_+,n_-)$ can also equivalently be expressed in terms of $(n_+,\mu)$  or $(G_{(4)}Q_m, \chi)$, where
$\chi$ is the orbifold Euler characteristic of the spindle given in \eqref{euler}.

Next it is convenient  to make a change of parameters which will allow us to parametrize the three-parameter family of solutions
in terms of three independent variables $(\bb,\bc,\bt)$. This is achieved via
\begin{align}
e \, =\,  \frac{\bb \bt}{\alpha^2 \bc}\, ,\quad  g \, = \,  \frac{\bt}{\alpha^2 \bc} \, ,\quad a \, = \, \frac{\bt}{\alpha} \, ,\quad\Longleftrightarrow\quad
\bb\, = \, \frac{e}{g}\, ,\quad \bc \, =\, \frac{a}{g\alpha}\, ,\quad \bt \, = \, a \alpha\, .
\label{magicchange}
 \end{align}
Notice that this is valid only if the rotation parameter $a$ is non-zero, and so we will continue with assuming
\begin{align}\label{anzero}
a \, \neq  \, 0\,.
\end{align}
The case $a=0$ should be examined separately, although we note that this non-rotating 
solution was studied in some detail in \cite{Ferrero:2020twa}. 
Recall also that in \eqref{signchoice} we initially took all parameters to be non-negative, and 
 moreover we are most interested in having a non-zero acceleration parameter $\alpha>0$. 
This then implies that $\bb\geq 0$, $\bc, \bt>0$ for physical solutions, although 
we shall shortly relax the requirement that all parameters are real. 
In what follows we shall therefore be careful to state 
what reality properties are being assumed when stating any given equation.

We may then proceed by expressing things in terms of the three parameters
$(\bb,\bc,\bt)$.
From the definition \eqref{Xidef} of $\Xi$, the first equation in \eqref{cond1} is equivalent to 
\begin{align}\label{genalpha}
\alpha^2\, = \, \frac{2 g\mu-1+a^2}{a^2+e^2+g^2}\, = \, \frac{\bt[2\bc\mu+\bt(\bc^2-1-\bb^2)]}{\bc^2(1+\bt^2)}\,.
\end{align}
 We next move to the thermodynamic quantities that do not depend explicitly on 
the horizon radius $r_+$. We find that the mass $M$ is given by 
\begin{align}
G_{(4)}M\, = \, \frac{1}{\alpha \bk}\frac{(\bc \bt+2\mu)(\bc-2\bt\mu)}{4 n_+\mu(1+\mu) \bc(1+\bt^2)}\, .
\end{align}
Using \eqref{kappa}
we then also compute
   \begin{align}\label{kapexp}
\alpha  \bk & \, =  \, \frac{\sqrt{\bt(\bc\bt +2\mu)(\bc-2\bt\mu)}}{\bc (1+\bt^2)}\, , 
\end{align}
where there is inherently a sign ambiguity in this equation due to the square root. 
For physical solutions we have 
$\bk>0$ and $M>0$, and hence $\bc > 2\bt \mu$, and we take the positive square root in 
\eqref{kapexp}. 
The mass $M$, angular momentum $J$ and electric charge $Q_e$ are given by
\begin{align}\label{MJQe}
G_{(4)}M& \, = \, 
\frac{G_{(4)}Q_m\sqrt{(2cG_{(4)}Q_m-\chi s)(2 c s G_{(4)}Q_m+\chi) }}{\chi \sqrt{s}}\,, \nn [1mm]
G_{(4)}J &\, = \, 
c \,(G_{(4)}Q_m)^2\,,         \nn [1mm]
G_{(4)}Q_e& \, = \, 
b \,(G_{(4)}Q_m)\, ,
\end{align}
where here we have replaced $(n_+,\mu)$ with $(\chi,G_{(4)}Q_m)$ using \eqref{euler},\eqref{gqmexps}.
Note that given the magnetic charge $Q_m$ (which is part of our fixed boundary data), the parameters $b$ and $c$ directly provide the electric charge $Q_e$ and the angular momentum $J$, respectively.

\subsection{Supersymmetry condition}\label{sec:SUSYcondition}

We now turn to imposing the second supersymmetry equation in \eqref{SUSY}. In the new variables 
this reads
\begin{align}\label{bpscondpaper}
\frac{\bt^2}{\bc^3\alpha^4}[-\bc(1-2 \bb \bt-\bt^2)+2 \bt\mu(1+\bb^2)]\, = \, 0\, .  
\end{align}
This is then immediately solved via 
\begin{align}\label{cbps}
\bc &\, = \, \frac{2(1+\bb^2)\bt\mu}{1-2\bb\bt-\bt^2}\, ,
\end{align}
where the denominator is assumed to be non-zero. 
The supersymmetry locus is thus
parametrized by the two parameters $(\bb, \bt)$. From \eqref{genalpha} we now have
\begin{align}\label{alphabps}
\alpha^2\, = \, \frac{4 \mu ^2 (1-\bb \bt)^2-\left(1-2 \bb \bt-\bt^2\right)^2}{4 \left(1+\bb^2\right) \left(1+\bt^2\right)\mu ^2 }\, ,
\end{align}
and from \eqref{kapexp} 
\begin{align}\label{alphakappa}
\alpha\bk\, = \, \frac{(\bb+\bt)(1-\bb\bt)}{(1+\bb^2)(1+\bt^2)}\, .
\end{align}
Focusing on the case where all parameters are non-negative, from $\bc>0$ 
we see from \eqref{cbps} that on the supersymmetry locus we must have $1-2\bb \bt-\bt^2>0$. We also have, trivially, $\bc+2\bb\mu>0$, and after substituting \eqref{cbps} we can also conclude that
$1-\bb \bt>0$ on the supersymmetry locus. Again, we shall shortly relax these conditions. 

Substituting the supersymmetry condition \eqref{cbps} into the mass $M$ in \eqref{MJQe}, one finds the perfect square
\begin{align}
\frac{(\bc \bt + 2\mu)(\bc - 2\bt \mu)}{\bt} \, = \, \left(\frac{2(\bb+\bt)(1-\bb \bt)\mu}{1-2\bb \bt -\bt^2}\right)^2\, .
\end{align}
Taking the square root that gives the quantity inside the bracket on the right hand side, 
one finds that the corresponding conserved charges in \eqref{MJQe} satisfy
the relation
\begin{align}\label{BPSrelation}
M\, = \, \frac{2}{\chi}J+Q_e\,,
\end{align}
where recall that $\chi=\chi(\SSigma)$ is the Euler characteristic of the surface $\SSigma$ introduced in~\eqref{euler}. This relation is expected to be a direct consequence of the supersymmetry algebra evaluated on the solution. 

\subsection{Horizon radius and extremal solutions}\label{sec:extreme}

A number of the thermodynamic quantities of interest in \eqref{allquantities}, \eqref{morequantities} depend 
on the horizon radius $r_+$, which recall is the largest root of the metric function $Q(r)$ in \eqref{PDfunctions}. 
On the other hand, by definition an extremal solution has a double root of $Q(r)$ at $r=r_+$. To 
examine this further it is convenient to define 
\begin{align}
r_+ \, \equiv\, \frac{\bt}{\alpha}\,\bR\, ,
\end{align}
and regard $\bR$ as a new parameter. Imposing the supersymmetry and regularity conditions from the previous subsections,
the condition $Q(r_+)=0$ reads
\begin{align}
\frac{(1+\bb^2)(1+\bt^2)\mu^2}{\left[\left(1-2 \bb \bt-\bt^2\right)^2-4 \mu ^2 (1-\bb \bt)^2\right]^2}\mathcal{Q}(\bR) \, = \ 0 \, , 
\end{align}
where we have introduced  
\begin{align}
\mathcal{Q}(\bR) & \, \equiv\, \left[\left(1-2 \bb \bt-\bt^2\right)^2+4 \mu ^2 (\bb+\bt)^2\right] \bt^4\bR^4 +4 \mu  \left(1+\bt^2\right) \left(1-2 \bb \bt-\bt^2\right) \bt^3 \bR^3  \nn
& \qquad +
2 \left[2 \mu ^2 \left(2 \bb^2 \bt^2+2 \bb \left(\bt^2-1\right) \bt+\bt^4+1\right)-\left(1-2 \bb \bt-\bt^2\right)^2\right] \bt^2\bR ^2 \nn
& \qquad-4 \mu    \left(1+\bt^2\right) \left(1-2 \bb \bt-\bt^2\right)\bt\bR +1\nn
& \qquad +\bt \left(4 \bb^2 \bt+4 \bb \bt^2+4 \mu ^2 \bt (1-\bb \bt)^2-4 \bb+\bt^3-2 \bt\right)\, .
\end{align}
Setting $\mathcal{Q}(\bR)=0$  is a quartic in $\bR$, as expected, but it is also a quadratic in $\bb$. Solving for the latter gives
\begin{align}\label{newbcomplex}
b\, = \, b_\pm\,  \equiv \,  \frac{2 \mu  \rho }{\rho ^2-1}+\frac{\left(1-s^2\pm2 \ii \mu  s\right)B(\rho, s)}{2s \left(\rho ^2-1\right)  \left(\rho ^2 s^2-1\mp\ii \mu s \left(\rho ^2+1\right) \right)}\, , 
\end{align}
where
\begin{align}\label{BRt}
B(\bR,\bt) \, \equiv\, (1-\bR^2)(1- \bR^2 \bt^2)+2 \mu  \left(1+\bR^2\right) \bR  \bt\,  .
\end{align}
From (\ref{newbcomplex}) we see that after imposing supersymmetry,  generically we cannot demand that $\bR,\bt,b$ are all real parameters, and as a consequence the physical charges and 
the entropy are complex quantities. From now on, in the remainder of this section, we will assume that 
\begin{align}
\bR\in \mathbb{R}\,,
\end{align} while we will allow the parameters $\bt$ and $b$ to be 
\emph{complex}\footnote{In the following analysis it is also possible to assume that $\bt \in \mathbb{R}$, but we shall not do so.}, and related by 
(\ref{newbcomplex}). 
Notice that  $\bb$ is real
 precisely when $B(\bR,\bt) =0$ (the parameter $\mu$ was introduced in \eqref{mudef}, and is necessarily real due to its
relation to the conical deficit angles on $\SSigma$.).

An extremal solution has a double root at $r=r_+$. We compute
\begin{align}
\mathcal{Q}'(\bR) \, = \, \frac{4\bt^2\mu ^2\left(1+\bt^2\right)^2 (\bR \mp \ii)  \left(\pm \ii \bR  \bt^2+\mu  (\bR \mp\ii) \bt-1\right)}{\left(1-\bR^2 \bt^2\pm \ii \mu  \left(1+\bR^2\right) \bt\right)^2}B(\bR,\bt)\, ,
\end{align}
where we have substituted for $\bb$ using \eqref{newbcomplex} with the $\pm$ signs 
correlated with that in \eqref{newbcomplex}. 
We are interested in a supersymmetric complexified solution that at extremality matches the Lorentzian supersymmetric and extremal solution, which is necessarily real. Hence at extremality the parameters $\bR$ and $\bt$ should both be real.
Setting $\mathcal{Q}'(\bR)=0$ to obtain a 
double root and demanding that both $\bR$ and $\bt\neq0$ are real implies either $\bR\bt=1$ and $\mu=1$ (which is 
not possible from \eqref{mudef})
or else $B(\bR,\bt)=0$, which gives
\begin{align}\label{bstar}
\bb_\star \, = \,  \frac{2\mu \bR}{\bR^2-1}\, ,
\end{align}
where we will denote the supersymmetric and extremal values of all quantities with 
a subscript $\star$. The equation \eqref{BRt} may then be viewed as a quadratic 
for $\bt$, and with $\bt=\bt_\star>0$ (from \eqref{magicchange}, \eqref{anzero}) the solution is
\begin{align}\label{sstar}
\bt_\star \, = \,  \frac{-\mu  \left(1+\bR^2\right)+\sqrt{\mu ^2 \left(1+\bR^2\right)^2+\left(\bR^2-1\right)^2}}{\bR \left(\bR^2-1\right)}\, ,
\end{align}
which is manifestly real.
From \eqref{rprange}, for the physical, real Lorentzian solution we have $0< \alpha r_+=\bt \bR<1$, and hence $0<\bt_\star < 1$ and $\rho>1$. 
We can thus parametrize the supersymmetric and extremal solutions, 
for fixed values of $\chi,G_{(4)}Q_m$ (or 
equivalently fixed $n_\pm$) by the parameter $\rho$, with
\begin{align}
\rho\, > \, 1\,.
\end{align}
Clearly, from \eqref{bstar} we have $\bb_\star>0$
and one can check from \eqref{cbps} that we also have $c=c_\star>0$.
We can also directly check that the extremality condition for supersymmetric solutions given in \eqref{extremal}, which  in the new variables\footnote{Note that if one just substitutes \eqref{cbps} into this extremality condition, one obtains the equivalent extremality condition $b^2(b^2+1)=c(c+2 b\mu)$.} reads
$\bb^2(1+\bb^2)^2\bt=\bc^2(1-\bb \bt)(\bb+\bt)$, is indeed satisfied after substituting \eqref{cbps}, \eqref{bstar} and \eqref{sstar}.

It is straightforward to now compute the thermodynamic quantities of section \ref{fstlaw} 
in the supersymmetric and extremal case as a function of $\rho$ for fixed $\chi,G_{(4)}Q_m$
and we find
\begin{align}\label{extremalcharges}
G_{(4)}M_\star &  \, = \,  
\frac{\bR \left(\sqrt{{\chi}^2 \left(\bR^2+1\right)^2+16 (G_{(4)}Q_m)^2 \left(\bR^2-1\right)^2}+{\chi} \left(\bR^2-1\right)\right)}{4 \left(\bR^2-1\right)^2}\,,\nn
G_{(4)}J_\star & \, = \ 
\frac{{\chi} \bR \left(\sqrt{{\chi}^2 \left(\bR^2+1\right)^2+16 (G_{(4)}Q_m)^2 \left(\bR^2-1\right)^2}-{\chi} \bR^2+{\chi}\right)}{8 \left(\bR^2-1\right)^2}\,,\nn
 \quad G_{(4)} (Q_e)_\star & \, = \, 
 \frac{{\chi} \bR}{2 \left(\bR^2-1\right)}\,,
\end{align} 
These satisfy\footnote{Notice that if we compare
the expression 
or $J_\star$ in \eqref{extremalcharges} and \eqref{MJQe} we deduce that $c=c_\star>0$, as noted above.}
the supersymmetry relation \eqref{BPSrelation}
\begin{equation}
M_\star \, = \, \frac{2}{\chi}J_\star + (Q_e)_\star\, ,
\label{veryEBPS}
\end{equation}
as well as the following non-linear relation between the charges \cite{Ferrero:2020twa}
\begin{align}\label{Jextremenice}
J_\star \,  = \, \frac{(Q_e)_\star}{4}\left(-\chi+\sqrt{\chi^2+(4G_{(4)})^2[(Q_e)_\star^2+Q_m^2]}\right)\, .
\end{align}

We may also compute the chemical potentials in the supersymmetric extremal case:
\be\label{extremalpotentials}
\begin{split}
T_\star \, = \, 0\, , \qquad \Omega_\star \, = \, \frac{2}{\chi}\,  , \qquad (\Phi_e)_\star \, = \, 1\, , \qquad (\Phi_m)_\star \ = \, \frac{1}{b_\star}\, .
\end{split}
\ee
The first equation, namely the black hole having zero temperature $T_\star=0$, was of course expected as it characterizes extremality. 
The supersymmetry relation \eqref{veryEBPS} may thus also be written as
\begin{align}
M_\star \, = \, (\Omega J)_\star + (\Phi_e Q_e)_\star\, .
\end{align}
We also find that the extremal value of Bekenstein-Hawking entropy is given by
\begin{align}\label{SBHextreme}
(\SBH)_\star&\, = \, 
\frac{\pi}{4G_{(4)}}\left( -{\chi}+
\frac{\sqrt{\chi^2 \left(\bR^2+1\right)^2+16(G_{(4)} Q_m)^2 \left(\bR^2-1\right)^2}}{\bR^2-1}\right)
\nn
&\, = \,  \frac{\pi}{G_{(4)}}\frac{J_\star}{(Q_e)_\star}\, .
\end{align}

The range of $\rho$ is given by $\rho\in (1,\infty)$. 
As ~$\rho\to \infty$ the supersymmetric and extremal solutions
approach the non-rotating black hole solutions. Although we have been considering the case $a>0$, the case of $a=0$
was considered in some detail in section 6 of \cite{Ferrero:2020twa}. In particular, it was shown there that the supersymmetric and extremal limit is then achieved when $e=0$. As a consequence these black holes have $J_\star=(Q_e)_\star=0$.
Furthermore, using (6.6) of \cite{Ferrero:2020twa} and the expression for the mass given in \eqref{allquantities} we find that
these black holes also have $M_\star=0$.  Taking the limit $\rho\to\infty$ in
\eqref{extremalcharges} precisely gives these values. Moreover, the  $\rho\to\infty$ limit
 of the first expression in \eqref{SBHextreme} gives the correct expression for the black hole entropy.

As $\rho \to 1$, from \eqref{sstar} we have $s_\star \to 0$ and hence $\alpha r_+\to 0$, which is excluded from our analysis, since we have focused on $\alpha>0$. It is worth noting however, that the correct thermodynamic 
expressions can be obtained for the non-accelerating, supersymmetric and extremal, electrically charged Kerr-Newman black holes by setting
$\chi=2$, $Q_m=0$ (obtained by formally setting $n_+=n_-=1$)
in \eqref{extremalcharges}-\eqref{SBHextreme}, along with setting $b_\star=\infty$, as suggested by \eqref{bstar}, 
so that $\Phi_m=0$.

\subsection{Complex supersymmetric locus}\label{sec:complexSUSY}

In the last subsection we have seen that we may parametrize solutions 
to the supersymmetry equations, for fixed 
$\chi,G_{(4)}Q_m$, in terms  
of the two variables $\bR\in \mathbb{R}$, $\bt\in \mathbb{C}$
and generically there are two branches of complex 
solutions with the parameter $\bb\in \mathbb{C}$ 
given by \eqref{newbcomplex}, and the parameter $c\in \mathbb{C}$ then determined 
from \eqref{cbps}. The parameter $\bb$ is real if and only if the solution 
is extremal, for which we then require
\begin{align}\label{ineqs}
\bt\, = \bt_\star > \, 0\, , \qquad \bR\, > \, 1\, ,
\end{align}
with $\bt_\star\in \mathbb{R}$, given in \eqref{sstar}.
In this section we study the family of complex supersymmetric but non-extremal
solutions for which the second inequality in 
\eqref{ineqs} also holds. 

For this complex supersymmetric locus we continue with the positive square root in~\eqref{kapexp}. We first find that
the supersymmetry relation \eqref{BPSrelation} between charges continues to hold. 
We also find that the chemical potentials satisfy
\begin{align}\label{SUSYchems}
\beta\left(1 + \frac{\chi}{2}\Omega -2\Phi_e\right) \, = \, \mp 2\pi \ii \, ,
\end{align}
where $\beta=1/T$, and in what follows the  signs are correlated with those 
of the two complex branches with $\bb=\bb_\pm$.  
We may then define the following complex chemical potentials:
\begin{align}
\omega \, \equiv\, \beta(\Omega -\Omega_\star)\, , \qquad \varphi\, \equiv \, \beta(\Phi_e- (\Phi_e)_\star)\, ,
\end{align}
where the extremal values of the chemical potentials $\Omega_\star$, $(\Phi_e)_\star$ are given in \eqref{extremalpotentials}.  
We find
\begin{align}
\label{omegasigma}
\omega \,& =\frac{4\pi(\bR\mp \ii)\bt}{\chi(-1\mp \ii \bR) \bt+4 G_{(4)}Q_m (\pm  \ii + \bR \bt^2)}\, ,\nn [1mm]
\varphi & \, = \, \pm \ii \pi +\frac{\chi}{4}\omega\, ,
\end{align}
and in particular the combination
\begin{align}\label{SUSYcon}
\varphi - \frac{\chi}{4}\omega \, = \, \pm \ii\pi
\end{align}
is independent of the parameters. 

Combining the supersymmetry relation \eqref{BPSrelation} with the first law \eqref{standardfirstlaw}, we arrive at the following supersymmetric form of the first law~\cite{Cassani:2019mms}
\begin{align}
\dd \SBH + \varphi\, \dd Q_e + \omega\, \dd J \,=\, 0\, .
\end{align}
Using \eqref{SUSYcon} this can equivalently be written in either of the following two forms:
\begin{align}
\dd (\SBH\pm\ii \pi Q_e) + \omega\, \dd \left(J +\frac{\chi}{4}Q_e\right)\, & =\, 0\, , \nn
\mbox{or} \qquad \dd \left(\SBH\mp \frac{4}{\chi}\ii \pi J\right) + \frac{4}{\chi} \varphi\,  \dd \left(J +\frac{\chi}{4}Q_e\right)\, & =\, 0\, . 
\label{Ssusyop1}
\end{align}
In particular, from either of the last two equations, it follows that $\omega=\omega (J +\frac{\chi}{4}Q_e)$, $\varphi=\varphi (J +\frac{\chi}{4}Q_e)$.
As we will see in the next section, the combination of charges $J +\frac{\chi}{4}Q_e$ commutes with the boundary supercharge.

Recall that in section \ref{fstlaw} we showed 
that the on-shell action $I=I(\beta,\Omega,\Phi_e)$ may be viewed as 
a function only of the chemical potentials. We have not 
found an explicit expression for this in general. However, for the 
complex supersymmetric solutions one can verify that we can write the action
as a complex function of the complex chemical potentials $\varphi,\omega$:
\begin{align}\label{SUSYaction}
I &\, = \,  -\SBH - \omega J - \varphi Q_e \, =\, \pm \frac{1}{2\ii G_{(4)}}\left[\frac{\varphi^2}{\omega} +  (G_{(4)}Q_m)^2\omega \right]\, .
\end{align}
Here all quantities are complex functions of $\bt$ and $\bR$:
\be
\begin{split}
\SBH & \, = \, \frac{2\pi Q_m (1+\bR^2)(1+\bb^2)\mu\bt^2}{(1-\bR^2\bt^2)(1-2\bb\bt-\bt^2)}\, , \quad 
J\, = \, G_{(4)}Q_m^2 \frac{2(1+\bb^2)\mu \bt}{1-2\bb \bt - \bt^2}\, , \quad 
Q_e\, = \, Q_m b\, ,
\end{split}
\ee
where one should substitute for $b$ given in (\ref{newbcomplex}). 

The formulae \eqref{SUSYchems}, \eqref{SUSYcon}, \eqref{SUSYaction} correctly reduce 
to those derived for the supersymmetric Kerr-Newman black holes in \cite{Cassani:2019mms}. 
Specifically, as noted above, one should (formally) set $n_-=n_+=1$ to obtain an $S^2$ horizon, which 
sets the Euler number $\chi=2$ and the magnetic charge $Q_m=0$ in the above formulae.

The extremal limit of this complex locus of supersymmetric solutions 
is obtained using \eqref{bstar}, \eqref{sstar}. In this limit we obtain the extremal ``starred" values for the conserved charges 
given in the previous subsection, all of which are real.\footnote{We prove a converse result in the next subsection.} We also obtain complex limiting expressions for
$(\omega_\star, \varphi_\star)$, still satisfying $\varphi_\star= \frac{\chi}{4}\omega_\star  \pm \ii\pi$,
and by substituting into \eqref{SUSYaction} we obtain an expression $I_*$ for the action, which is complex.
When we take the extremal limit we recover the real supersymmetric and extremal 
solutions of interest and it may seem strange that the action is complex in this limit. However, we should recall that
the action is not defined for the real extremal solutions since this involves, in the Euclidean section, taking $\beta\to\infty$. Thus 
$I_\star$ can be viewed as a definition.
Of more interest is that after a Legendre transform we can obtain an expression
for the black hole entropy along the supersymmetric and complex locus, which recovers the entropy of the
real supersymmetric and extremal black holes, as we discuss next.

\subsection{Legendre transform}\label{sec:legendre}

Let us start from the supersymmetric on-shell action from the last subsection:
\begin{align}\label{susyaction}
I \, = \,  \pm\frac{1}{2\ii G_{(4)} }\left[\frac{\varphi^2}{\omega} + (G_{(4)}Q_m)^2\omega \right]\, .
\end{align}
This is minus the logarithm of a supersymmetric grand-canonical partition function, depending on $\varphi$, $\omega$, where the electrostatic and rotational chemical potentials are subject to the constraint
\be\label{susy_constraint}
\varphi - \frac{\chi}{4}\, \omega \, = \, \pm\pi \ii\,.
\ee
The action also depends on the magnetic charge $Q_m$, that is always held fixed in the problem under study.
The entropy, for given $(G_{(4)}Q_m,\chi)$, is given by the Legendre transform 
\begin{align}\label{defssss}
\SBH(J,Q_e) & \,=\, {\rm ext}_{\{\omega,\varphi,\Lambda\}}\Big[- I(\omega, \varphi) - \omega J - \varphi Q_e- \Lambda \left(\varphi - \frac{\chi}{4} \omega \mp  \pi \ii\right) \Big]\,,
\end{align}
where $\Lambda$ is a Lagrange multiplier enforcing the constraint \eqref{susy_constraint}; here we are following the method of appendix B in \cite{Cabo-Bizet:2018ehj}. The entropy is the logarithm of the micro-canonical partition function and thus depends on the charges, that is $J$ and $Q_e$ (for given $(G_{(4)}Q_m,\chi)$, both of which are fixed by
the spindle horizon).

The extremization equations are given by 
\be\label{exteqs}
-\frac{\partial I}{\partial\omega} \,=\, J - \frac{\chi}{4}\,\Lambda\,,\qquad\qquad -\frac{\partial I}{\partial\varphi} \, = \, Q_e + \Lambda\,,
\ee
together  with the constraint \eqref{susy_constraint}. 
Substituting for the 
 derivatives using \eqref{susyaction} we  deduce
\begin{align}\label{lambdetc}
\Lambda&\, = \, -Q_e\pm \frac{\ii\chi}{4 G_{(4)}}+\frac{\ii \eta}{4 G_{(4)}}{\sqrt{\chi^2 + (4G_{(4)}Q_m)^2 \pm 8\ii G_{(4)} (\chi Q_e +4 J) }}\, , \nn[1mm]
\omega &\, = \,  \frac{4 \pi \ii  \eta}{\sqrt{\chi^2 + (4 G_{(4)}Q_m)^2 \pm 8\ii G_{(4)}(\chi Q_e +4 J) }}\, ,\nn [1mm]
\varphi &\, = \, \frac{\chi}{4}\omega\pm  \pi \ii\, , 
\end{align}
where $\eta=\pm 1$. 
At the extremum we then have 
\be\label{Legendre_transform_result}
\SBH  \, =\, \pm\pi \ii \Lambda\,,
\ee
with $\Lambda$ as in \eqref{lambdetc}. To see this, note that since $I$ is homogeneous of degree one in $\varphi,\omega$, we have
$I=\varphi\frac{\partial I}{\partial\varphi}+\omega\frac{\partial I}{\partial\omega}$ and then one can use 
\eqref{exteqs} in \eqref{defssss}. Note that from the expression for $\omega$ in (\ref{lambdetc}) it is clear that this depends on the charges that are being varied in the extremization, only through the combination 
$J +\frac{\chi}{4}Q_e$, consistent  with~\eqref{Ssusyop1}. 

If we now assume that $\SBH$, $Q_e$ and $J$ are real then we recover the supersymmetric extremal limit, as we now argue. We first note that from \eqref{lambdetc}, \eqref{Legendre_transform_result} we can deduce 
\begin{align}\label{extafterleg}
J \, = \,   \frac{Q_e}{4}\left(-\chi+\sqrt{\chi^2+(4G_{(4)})^2 \left(Q_e^2+Q_m^2\right)}\right)\, , \qquad \SBH\, = \, \frac{\pi}{G_{(4)}}\frac{J}{Q_e}\,, 
\end{align}
which are  precisely the extremal values \eqref{Jextremenice}, \eqref{SBHextreme}, where we have chosen the sign to
ensure that $J>0$. To complete the argument, we next observe from \eqref{MJQe} that $b,c$ must be real and $s$ is constrained via \eqref{cbps}. Proceeding, the second condition in \eqref{extafterleg}
expressed in terms of the parameters implies $b = \frac{1-\bR^2 \bt^2}{\bt(1+\bR^2)}$ and after substituting this into
the first condition in \eqref{extafterleg} implies
$[ (1-\bR^2)(1- \bR^2 \bt^2)-2 \mu  \left(1+\bR^2\right) \bR  \bt] B(\bR,\bt)=0$.
We now find that with $\bt>0$, $\bR>1$ the only possibility is the extremal solution we found in section \ref{sec:extreme} with, in particular, $B(\bR,\bt)=0$.

\section{Euclidean supersymmetric action from a fixed point formula}\label{sec:boundary}

In this section we will recover our expression for the on-shell Euclidean supersymmetric action \eqref{susyaction}
using a general fixed point formula for gravitational solutions 
 that was presented in  \cite{BenettiGenolini:2019jdz}. In order to do this, we have to compute a canonical Killing vector 
 possessed by the family of supersymmetric solutions, which we will extract, slightly indirectly, from a corresponding supersymmetric Killing vector of the boundary geometry. To obtain the latter we first obtain the boundary Killing spinor,
 which solves the conformal Killing spinor equation, and then use it to construct the boundary Killing vector as a suitable bilinear. By continuity, this boundary Killing vector can then be extended into the bulk.
  
AdS/CFT implies that we should be able to identify the on-shell 
action \eqref{susyaction} with minus the logarithm of an appropriate supersymmetric index 
of the boundary field theory.
For such a comparison with field theory, the uplift to $D=11$, briefly summarized at the end of section \ref{regresults}, is certainly important. Furthermore, we anticipate that the boundary Killing vector will play a key role in
a direct evaluation of the corresponding supersymmetric partition function of the dual field theories.
 
\subsection{Boundary Killing spinor and Killing vector}

In order to perform the computations of the present section, we have found it technically more convenient to start by working in Lorentzian signature, assuming that all the parameters 
are real. In practice, this means that we treat the Killing spinors as spinors in Lorentzian signature, 
with the usual rules for charge conjugation. 
We then Wick rotate and analytically continue any results of interest to complex parameters
at the end, in 
particular once we have computed the Killing vector bilinear.

We begin by introducing the following orthonormal frame for the boundary 
metric~\eqref{metric_bdry}: 
\begin{equation}
\begin{split}
e^0&\, = \, \sqrt{\tilde{P}} \left(\frac{1}{\bk}\diff t - a f \diff\phi\right)\, , \qquad 
e^1 \, = \, -\sqrt{F}\, \diff x\, , \qquad e^2 \, =\, \sqrt{G}\, \diff\phi\, ,
\end{split}
\end{equation}
where as in \eqref{xvar} it is convenient to use the variable
\begin{align}
x \, \equiv \, \cos\theta\, .
\end{align}
In a slight abuse of notation we write the metric functions \eqref{PDfunctions}, \eqref{tildePdef} as
\begin{align}\label{bdryone}
P & \, = \, P(x)\, \equiv\, 1-2\alpha m x + \left(\alpha^2(a^2+e^2+g^2)-a^2\right)x^2\, ,\nn
\tilde{P} & \,  = \, \tilde{P}(x)\, \equiv \, 1-\alpha^2 (1-x^2)P(x)\, ,
\end{align}
and we have introduced the functions
\begin{align}\label{bdrytwo}
f & \, = \, f(x)\, \equiv\, \frac{1-\alpha^2P(x)}{\tilde{P}(x)}(1-x^2)\, ,\nn[1mm]
F& \, = \, F(x)\, \equiv\, \frac{(1+a^2\alpha^2 x^4)^2}{P(x)(\tilde{P}(x)+a^2\alpha^2x^4)(1-x^2)}\, ,\nn[1mm]
G& \, = \, G(x)\, \equiv\, P(x)\left(1+\frac{a^2\alpha^2x^4}{\tilde{P}(x)}\right)(1-x^2)\, .
\end{align}
The boundary gauge field \eqref{gfdexpbdy} is 
\begin{align}
A_{\mathrm{bdy}} \, = \, A^{\mathrm{bdy}}_t\, \diff t + A^{\mathrm{bdy}}_\phi\, \diff\phi\, ,
\end{align}
where
\begin{align}\label{someayexps}
A^{\mathrm{bdy}}_t & \, \equiv \,  -\frac{\alpha \,x}{\bk(1+a^2\alpha^2x^4)}\,(e - g a\alpha x^2)\, , \nn
A^{\mathrm{bdy}}_\phi &  \, \equiv \,   -\frac{x}{1+a^2\alpha^2x^4}\left[g+ga^2\alpha^2 x^2-ea \alpha (1-x^2)\right]\, .
\end{align}

We eliminate the parameter $m$ by using the first supersymmetry condition $m=g/\alpha$, then change  variables from $(e,g,a)$ to $(\bb,\bc,\bt)$ via \eqref{magicchange}, and finally impose the second supersymmetry condition by imposing the equation 
\eqref{cbps} for the parameter~$\bc$.
 As we explained at the beginning of the section, this leads to a two-parameter family of real supersymmetric Lorentzian solutions, parametrized by the real constants $\bb, \bt$, together with the parameters $(G_{(4)}Q_m,\chi)$, or equivalently $n_\pm$, which determine the conical deficits of the spindle horizon surface $\SSigma$. 
 The parameter $\alpha$ is given in terms of $(\bb,\bt,\mu)$ by \eqref{alphabps}, where $\mu$ is defined in \eqref{mudef},  
 and $\alpha\bk$ is given in~\eqref{alphakappa}. 

The bulk Killing spinor equation \eqref{KSE} for minimal $D=4$, $\mathcal{N}=2$ gauged supergravity induces \cite{Hristov:2013spa,Cassani:2012ri} the 
following conformal Killing spinor equation (CKSE) on the conformal boundary 
\begin{align}
\nabla_i\zeta\, = \, \frac{1}{3}\gammathree_i \gammathree^j\nabla_j \zeta\, ,
\end{align}
where the covariant derivative is $\nabla_i\equiv\partial_i + \frac{1}{4}\omega_i^{\ jk}\gammathree_{jk} - \ii A^{\mathrm{bdy}}_i$.
We use the gamma matrix conventions of \cite{Ferrero:2020twa}, namely
$\gammathree_0=\ii \sigma^1$, 
$\gammathree_1=\sigma^2$, $\gammathree_2=\sigma^3$, in terms of Pauli matrices $\sigma^a$, with $\gammathree^0\gammathree^1\gammathree^2=+1$. 
After a lengthy calculation 
we find the solution\footnote{Recall that for these real Lorentzian supersymmetric solutions, discussed in section \ref{sec:SUSYcondition}, we had 
$1-\bb\bt>0$, in particular due to equation \eqref{alphakappa}. More generally ${\rm sign}(1-b s)$ appears 
as a factor in the lower component of \eqref{form_Killingspinor}.}
\begin{align}\label{form_Killingspinor}
\zeta \,&=\, \rme^{-\ii\,(\mu_1 t + \mu_2 \phi)}\begin{pmatrix} \sps\\  -\sps^*\end{pmatrix} \,, 
\end{align}
where $\sps=\sps(x)$ is a complex function,  and $\mu_1,\mu_2$ are real constants given by
\begin{align}
\mu_1 & \, = \, \frac{1+a^2\alpha^2}{2(1- a\alpha e/g)}\,=\, \frac{1+\bt^2}{2(1-\bb \bt)}\, , \nn [6pt]
\mu_2 & \, = \,  g\mu \, = \, \frac{2\mu^2(1+\bt^2)(1-2\bb\bt-\bt^2)}{4\mu^2(1-\bb\bt)^2-(1-2\bb\bt-\bt^2)^2}\, .
\end{align}
The complex function $\sps(x)$ is
\be
\sps \,  =\,  \sqrt{\frac{1}{2}\left[  \frac{(1-2\bb \bt-\bt^2)}{2\mu}(1-x^2) + (1+\bt^2) x + \ii\, \alpha (\bb + \bt) \sqrt{G}\right]}\,,
\ee
or, in terms of just the original variables,
\be
\sps \, = \,  \sqrt{\frac{1}{2}\left[  \frac{\alpha^2}{g}(e^2+g^2)(1-x^2) +(1+a^2\alpha^2) x + \ii\,\alpha\left(\frac{e}{g} + a \alpha\right) \sqrt{G}\right]}\,,
\ee
where the metric parameters $(e,g,a,\alpha)$ should be substituted for their supersymmetric values, 
parametrized by $(b,s,\mu)$, as described just below \eqref{someayexps}.

Next defining the usual Dirac adjoint of the Lorentzian spinor $\zeta$ as 
\begin{align}\label{barzeta}
\bar\zeta \,&\equiv \, \zeta^\dagger\gammathree_0 \,=\, -\ii\, \rme^{\ii(\mu_1 t + \mu_2 \phi)} \big(\; \sps\  \  -\sps^* \; \big)\,,
\end{align}
we may introduce the vector bilinear 
\be
K \,=\, K^\mu \partial_\mu \,\equiv \, \bar\zeta \gammathree^\mu \zeta\, \partial_\mu\, .
\ee
Substituting in for the metric functions and the solution for the Killing spinor, remarkably we find the simple expression 
\begin{align}\label{susyvector_Lorentzian}
K \,&=\, \kappa(1+\bt^2)\partial_t + \alpha(\bb+\bt)\partial_\phi\nn
\,&=\, \kappa(1+a^2\alpha^2)\partial_t + \alpha\left(\frac{e}{g}+a\alpha\right)\partial_\phi\,.
\end{align}
This is manifestly a Killing vector of the boundary metric, preserving also the gauge field.

\subsection{Analytic continuation and fixed point formula}\label{sec:globalSUSY}

Recall that in section \ref{sec:BHentropy} we introduced the primed coordinates \eqref{primedcoordinates}
\begin{align}\label{coc}
t\, =\, t',\qquad  \phi\, =\, \phi'+\Omega_H \frac{\Delta \phi}{2\pi} t'\, ,
\end{align}
where $\Omega_H$ is the angular velocity of the horizon \eqref{angvelh}. The null 
generator of the latter  is then 
 $\nullH=\partial_{t'}$, and furthermore these are the natural coordinates to
show that the Euclidean metric is regular at the horizon. This involves the Wick rotation
\begin{align}\label{wicky}
t' \, = \, -\ii \tau'~,
\end{align}
where recall that the Euclidean black hole then has topology $\R^2\times \SSigma$. It is then  convenient to introduce 
$2\pi$-period angular coordinates
\begin{align}
\psi \, \equiv\, \frac{2\pi}{\beta}\tau'\, , \qquad \ssigma\, \equiv\, \frac{2\pi}{\Delta\phi}\phi' \, .
\end{align}
Here $\psi$ is a polar angular coordinate on the $\R^2$ normal to the horizon, while 
$\ssigma$ is an azimuthal coordinate on the spindle horizon $\SSigma$. 

Having performed this Wick rotation to a Euclidean signature solution 
with topology $\R^2\times\SSigma$, we would next like to analytically continue 
to complex values of the parameters, as discussed in section~\ref{sec:SUSYBHs}. 
In principle we could have started directly with these complex 
solutions, with the Lorentzian presentation above adapted with relatively little change. 
In that case the boundary spinors $\zeta$ and $(\bar\zeta)^T$ should be regarded as independent spinors with opposite charge under $A_{\rm bdy}$, each solving the corresponding conformal Killing spinor equation. 
The supersymmetric Killing vector is still given by \eqref{susyvector_Lorentzian}, with $t \to -\ii \tau$ and with the parameters appearing there now taking complex values.  Since the Killing vector is complex, notice that this background is not immediately included in the classification of~\cite{Closset:2012ru} (not even when the spindle is a regular two-sphere). 
However, rather than taking this approach, we instead simply analytically continue the 
real solutions we have constructed. 
It is straightforward to compute the Killing vector field \eqref{susyvector_Lorentzian} 
under the above change of coordinates \eqref{coc} and Wick rotation \eqref{wicky}.

With this perspective in mind and imposing
also the complex supersymmetric locus, with $b=b_\pm$ given by \eqref{newbcomplex}, 
we find that 
\begin{align}\label{KEuclid}
K\, = \, 
\mathcal{N}\left[\partial_\psi \mp \ii \left(\frac{\omega}{2\pi}\right) \partial_\ssigma\right]\, .  
\end{align}
Here $\omega=\omega_\pm$ is given by \eqref{omegasigma}, and the 
overall normalization factor is $\mathcal{N}=2\pi \ii \kappa(1+\bt^2)/{\beta}$, although the latter may be 
rescaled by simply rescaling the Killing spinor $\zeta$ by an overall constant. Notice that \eqref{KEuclid} is generically complex.

Geometrically 
$\partial_\psi$ and $\partial_\ssigma$ rotate the two factors in the product $\R^2\times\SSigma$ 
with weight one, and the angular velocity $\ii\omega$ then appears as the relative 
weight between these generators in the supersymmetric Killing vector in \eqref{KEuclid} in a natural way. 
We note that in reference \cite{BenettiGenolini:2019jdz} a general 
formula for the holographically renormalized action $I$ of 
Euclidean supersymmetric asymptotically locally AdS solutions 
of minimal gauged supergravity was presented. This formula depends only 
on the fixed points of the supersymmetric Killing vector field 
$K$ in the bulk. Although we have only 
computed the restriction of this Killing vector to the conformal boundary, 
since $\partial_t$ and $\partial_\phi$ are generically the only 
Killing vectors of the solution, it follows by continuity that 
\eqref{KEuclid} must coincide with the bulk Killing vector. 
The formula in \cite{BenettiGenolini:2019jdz} was derived 
for real Euclidean solutions, although since our 
complex solutions arise from an analytic continuation
of real Lorentzian supersymmetric solutions, 
we expect the result of \cite{BenettiGenolini:2019jdz} to still hold. 
For non-zero $\omega$ the fixed points of $K$ 
in \eqref{KEuclid} are at the north and south poles 
of the spindle $\SSigma$ located at the black hole horizon,
which is at the origin of the $\R^2$ factor. 
Writing 
\begin{align}\label{Ka1a2}
K \, = \, a_1\partial_\psi + a_2\partial_\ssigma\, , \qquad 
\mbox{where}\ \quad a_1\, = \, \mathcal{N}\, , \quad  a_2\, = \, \mp \ii \, \mathcal{N} \left(\frac{\omega}{2\pi}\right) \, ,
\end{align}
the general formula for the Euclidean action in \cite{BenettiGenolini:2019jdz}
reads
\begin{align}\label{fixedpointI}
I\, = \, \left[\frac{1}{n_+}\frac{\left(\frac{a_2}{n_+}+a_1\right)^2}{4a_1\frac{a_2}{n_+}}+\frac{1}{n_-}\frac{\left(\frac{a_2}{n_-}+a_1\right)^2}{4a_1\frac{a_2}{n_-}}\right]\frac{\pi}{2G_{(4)}}\, .
\end{align}
Notice here that $\partial_\psi$ has weight 1 on the normal $\R^2$ to the horizon, while 
$\partial_\ssigma$ has weights $1/n_\pm$ on the tangent spaces to the 
poles of spindle horizon $\SSigma$, which are $\R^2/\Z_{n_\pm}$, respectively. 
The overall factors of $1/n_\pm$ in each of the two  terms in \eqref{fixedpointI} similarly arise because of the orbifold 
singularities. Substituting in for the values of $a_1$, $a_2$ in \eqref{Ka1a2}, we precisely 
recover our Euclidean supersymmetric action \eqref{susyaction} 
from the fixed point formula \eqref{fixedpointI}.

We can also examine the phase of the Killing spinor \eqref{form_Killingspinor} in the above Euclidean 
continuation. As discussed already in section \ref{sec:BHentropy}, our original gauge 
field is not regular at the black hole horizon, and indeed in general there is no gauge 
in which this is the case, due to the magnetic flux through $\SSigma$. However,  in the 
case when $g=0$ the gauge field becomes completely regular if we make a 
gauge transformation of the form  \eqref{gaugets} with 
$\alpha_1=(e r_+)/[\bk (a^2+r_+^2)]$. When $g\neq 0$ this gauge transformation 
makes the gauge field regular everywhere, except at the poles of the horizon $\SSigma$.
In this gauge, and evaluating on the complex supersymmetric locus with $b=b_\pm$ given by \eqref{newbcomplex}
we find that the Killing spinor \eqref{form_Killingspinor} reads 
\begin{align}\label{zetaanti}
\zeta \, = \, \ex^{\frac{\ii}{2}\left(\pm \psi - \frac{\chi}{2}\ssigma\right)}\begin{pmatrix} \sps\\ -\sps^*\end{pmatrix} \, .
\end{align}
In particular we see that this is anti-periodic around the Euclidean time circle $\psi$, which has period $2\pi$. 
This is  necessary in order that the bulk Killing spinor is smooth at the horizon, as only 
the anti-periodic spin structure extends smoothly to the origin of $\R^2$. 
Of course this is a slightly delicate statement, as these are complex solutions. 
However, \emph{before} imposing the complex supersymmetric locus we have 
real, non-extremal Euclidean solutions. The thermal circle 
and radial direction together form a ``cigar'' geometry, and any 
spinor field must be anti-periodic around the thermal circle in order to be non-singular at the horizon
(in a gauge that is regular at the tip of the cigar, for a fixed point on the spindle). 
We then complexified the solutions and imposed supersymmetry, 
and \eqref{zetaanti} shows that the resulting Killing spinor 
is anti-periodic. This is a very reasonable regularity 
condition to maintain for these complex solutions, where {\it a priori}
the precise regularity conditions one wants to impose are perhaps not clear.

Finally, notice that the Killing spinor $\zeta$ in \eqref{zetaanti} has charge
$-\chi/4$ under $J$, which is generated by $\partial_\sigma$. It also 
has $R$-charge 1 under the $R$-symmetry gauge field $A_\mathrm{bdy}$, as one sees from 
the Killing spinor equation \eqref{KSE}. It follows that $\zeta$ has 
charge zero under $J+\frac{\chi}{4}Q_e=-\frac{\chi}{4}+\frac{\chi}{4}=0$, and the 
corresponding supercharge $\mathcal{Q}$ in field theory 
should then commute with the operator $J+\frac{\chi}{4}Q_e$.
It is also interesting to point out that $J+\frac{\chi}{4}Q_e$ is precisely the same quantity as
$J_{AdS_2}$, the angular momentum of the near horizon solution that was defined in 
 \cite{Ferrero:2020twa}. In particular, recalling that the angular momentum is gauge dependent, $J_{AdS_2}$ is defined 
with a gauge field that is invariant under the symmetries of $AdS_2$. The new observation here is that the
relation $J_{AdS_2}=J+\frac{\chi}{4}Q_e$ of \cite{Ferrero:2020twa} shows that the charge $J_{AdS_2}$ is the one that
commutes with the field theory supercharge.

\section{Discussion}\label{sec:disc}

Using holographic techniques we have carried out a detailed analysis of the thermodynamics for
a general class of accelerating black hole solutions of $D=4$ minimal gauged supergravity. The black holes are rotating and carry electric and magnetic charges and lie within the family\footnote{We did not consider the possibility of NUT charge, which is included in the solutions of \cite{Plebanski:1976gy}, since we did not want to include closed timelike curves.}  
constructed in \cite{Plebanski:1976gy}. Of particular interest is that by taking the horizon to be a spindle and suitably
constraining the parameters, one can uplift on regular Sasaki-Einstein manifolds to obtain $D=11$ solutions that are free from
conical singularities \cite{Ferrero:2020twa}. In particular, this construction requires that the magnetic charge, which is specified by the spindle data,
is non-vanishing. Furthermore, the $D=11$ solutions preserve supersymmetry when 
the $D=4$ solutions do.

After holographically defining a set of conserved charges, we presented a first law which generalizes the result of \cite{Anabalon:2018qfv} to include magnetic charge. To obtain the first law, 
as in \cite{Anabalon:2018qfv}, 
when the acceleration is non-vanishing, $\alpha\ne0$, it was necessary to choose a specific, constant scaling, 
$\kappa$, of the time coordinate with, crucially, $\kappa$  depending on the parameters of the solution. While a constant and parameter independent scaling of the time coordinate corresponds to a simple scaling of dimensionful quantities in the dual field theory, the full significance of the parameter dependence of $\kappa$ for the accelerating black holes
warrants further study. In \cite{Anabalon:2018qfv} some justification of the specific form of $\kappa$ was given by considering the limit of vanishing black hole mass and string tensions, when the spacetime is then $AdS_4$ spacetime written in Rindler coordinates. However, these considerations do not fully fix $\kappa$ and, in fact, we found the precise form of
$\kappa$ that gives the first law by trial and error. 
It would certainly be interesting to have a better understanding of $\kappa$; the fact that the conformal boundary is not conformally flat, along with the fact that the variations entering the first law change the local conformal class of the boundary, appear to be significant features. 
It would also be interesting to make a direct connection with the approach of \cite{Papadimitriou:2005ii}.

We have also studied in some detail the one-parameter family of supersymmetric and extremal black holes,
which is the class where we hope to make precise contact with the dual field theory in future work, as we discuss below. Adopting the approach of~\cite{Cabo-Bizet:2018ehj,Cassani:2019mms}, 
we relaxed the extremality condition and analytically continued some of the parameters appearing in the black hole
solutions so as to identify a locus of complex supersymmetric solutions. On this complex locus we showed that the on-shell Euclidean action can be expressed as a function of complex rotational and electric chemical potentials which satisfy a constraint, and, moreover, the black hole entropy of the supersymmetric and extremal black holes can be recovered via a Legendre transformation and then imposing a reality constraint.
The expression of this supersymmetric on-shell action generalizes the one given in \cite{Cassani:2019mms,Bobev:2019zmz} to the accelerating case. It seems likely that this complex action
can be suitably identified with minus the logarithm of the supersymmetric partition function of the dual field theory.
From the gravitational point of view, we are considering a class of complex saddle points of the gravitational path integral.
In our formulation the underlying Euclidean manifold is real, but we are considering complex metrics and spinors that are obtained by analytic continuation of the parameters appearing in the solutions. Another interesting topic for future research is to elucidate more intrinsic criteria for determining which complex metrics should be considered along with which precise notions of spinors and supersymmetry one should use.

The results of our paper imply that the geometry of the conformal boundary provides a supersymmetric background where one can define three-dimensional ${\cal N}=2$ 
supersymmetric field  theories. It is then natural to conjecture that the supersymmetric partition function in this background will define a generalized index of the field theory. 
Specifically, after Wick rotating and compactifying the time direction on a circle, the background is $S^1\times \mathbb{\Sigma }$, together with a
background $R$-symmetry gauge field $A_R\equiv A_{\mathrm{bdy}}$, such that $\tfrac{1}{2\pi}\int_\mathbb{\Sigma}\diff A_R =  \frac{n_--n_+}{2n_+n_-}$. On general grounds, we expect this partition function to take the form
\begin{align}\label{conjindex}
Z(n_+,n_-,\varphi,\omega)_{S^1\times \mathbb{\Sigma}} \ & = \ \mathrm{Tr}_\mathrm{twist} \, \mathrm{e}^{-\beta \{ \mathcal{Q},\mathcal{\bar Q}\}}  \mathrm{e}^{\omega J  + \varphi Q_e} \nn [6pt]
& = \ \mathrm{Tr}_\mathrm{twist} \, \ex^{\pm \ii \pi Q_e}  \mathrm{e}^{-\beta \{ \mathcal{Q},\mathcal{\bar Q}\}   + \omega(J+\frac{\chi}{4}Q_e) }\, ,
\end{align}
where to go from the first to the second line we used the  constraint 
 \begin{align}
\varphi - \frac{\chi}{4}\omega \, = \, \pm \ii\pi\, .
\end{align}
Here $\mathcal{Q}$ is the supercharge of the theory compactified  on $\mathbb{\Sigma}$, and recall that the combination  $J +  \tfrac{\chi}{4}Q_e$ is the operator commuting with $\mathcal{Q}$, as discussed 
at the end of section~\ref{sec:boundary}. 
The subscript ``twist'' on the trace indicates  the twisting that we are performing  is \emph{different} from the topological twist for which a corollary is $\tfrac{1}{2\pi}\int_\mathbb{\Sigma}\diff A_R =  \frac{\chi}{2}=\frac{n_-+n_+}{2n_+n_-}$, whereas we have $\tfrac{1}{2\pi}\int_\mathbb{\Sigma}\diff A_R = 2 G_{(4)}Q_m=\frac{n_--n_+}{2n_+n_-}$. 
Notice that since $Q_e$ is the $R$-charge, the bosons and fermions within a multiplet have 
$Q_e$ values differing by 1, and 
the expression \eqref{conjindex} is hence indeed an index.\footnote{Furthermore, notice that 
shifting $\omega \equiv \tilde\omega\mp \frac{4\pi \ii}{\chi}$, the expression \eqref{conjindex} for the partition function becomes 
$\mathrm{Tr}_\mathrm{twist}\, \ex^{\mp\frac{4\pi \ii}{\chi}J} \mathrm{e}^{-\beta \{ \mathcal{Q},\mathcal{\bar Q}\}   + \tilde\omega(J+\frac{\chi}{4}Q_e) }$. Setting $n_-=n_+=1$, so that $\chi=2$, the first factor is $ \ex^{\mp 2\pi \ii J}=(-1)^F$, and since 
$Q_m=0$ there is no magnetic flux and hence no twist. This is then the same index computed in \cite{Nian:2019pxj}.}
 In the large $N$ limit this should reproduce
the entropy function~(\ref{susyaction}) for $I(\varphi,\omega)$,  but it will be of independent interest as an exact field theory object.

It is interesting to note that the  expression (\ref{susyaction}) for $I(\varphi,\omega)$ makes sense even in the non-rotating limit, that corresponds to $\rho\to\infty$ in our parametrization of the supersymmetric bulk solution. Indeed in this limit, before imposing
extremality, there is still a (complex) one-parameter family of solutions, parametrized by $s$. This implies  that the above index in fact will  also capture
 the entropy of the static accelerating black holes. 
In order to define this index, the main technical issue that will need to be addressed is what are the appropriate choices of 
boundary conditions on the fields at the orbifold singularities.

In \cite{Ferrero:2020twa} it was shown that uplifting the accelerating black hole solutions on 
Sasaki-Einstein spaces in the regular class can give rise to regular solutions in $D=11$. There are other ways to uplift the $D=4$ solutions
to $D=11$ or $D=10$ \cite{Gauntlett:2007ma,Donos:2010ax,Guarino:2015vca}
(in fact, locally this is possible whenever there is a supersymmetric $AdS_4$ solution  \cite{Gauntlett:2007ma,Cassani:2019vcl}), but generically they will be singular. 
It would be interesting to explore these uplifted solutions in more detail, and investigate whether or not it is still possible to make precise comparisons with the associated dual field theories. In fact, it has recently been shown \cite{Ferrero:2021wvk} that wrapping M5-branes on a spindle and then uplifting to $D=11$ on a four-sphere gives rise to solutions with orbifold singularities and yet a holographic computation of the central charge of the $d=4$ SCFT was found to precisely agree with a field theory computation.


\subsection*{Acknowledgments}
We thank Ioannis Papadimitriou and Kostas Skenderis for helpful discussions.
This work was supported by STFC grants ST/T000791/1 and ST/T000864/1.
 JPG is supported as a Visiting Fellow at the Perimeter Institute. 

\appendix

\section{Supersymmetric boundary in canonical form}\label{appa}

For supersymmetric solutions, the boundary metric and gauge field, given in \eqref{metric_bdry}, \eqref{gfdexpbdy}, can be recast in a canonical form. Specifically, we show that they lie within
the time-like class of three-dimensional supersymmetric rigid Lorentzian geometries which solve the charged conformal Killing spinor equation,
as studied in \cite{Hristov:2013spa}. This class is associated with the 
Killing vector bilinear, $K$, being time-like. Choosing coordinates such that 
\begin{align}\label{K_canonical}
K & \, = \, \frac{\de}{\de \tilde t}\, ,
\end{align}
 the three-dimensional metric and gauge field can be written
\begin{align}\label{3d_susy_background}
\dd s^2_{\rm can} \,&=\, \Upsilon^2\left[- (\dd\tilde t + \tilde\omega)^2 + \dd s^2_2\right]\,,\nn
A_{\rm can} \,& =\, -f\,(\dd\tilde t + \tilde\omega) + A_{(2)} \,,
\end{align}
where $\Upsilon$, $f$ are functions and $\tilde\omega, A_{(2)}$ are one-forms on the locally defined 2d base transverse to $K$ with metric $\diff s^2_2$, and thus all independent of $\tilde t$. 
 These quantities have to satisfy \cite{Hristov:2013spa}
\begin{align}\label{eq_omega}
\dd\tilde\omega &\, = \, 2f\,{\rm vol}_2\,,\nn
\dd A_{(2)} &\, = \, -\frac{1}{4} R_2\,{\rm vol_2}\,,
\end{align}
where ${\rm vol}_2$ and $R_2$ are the volume form and the Ricci scalar for $\dd s^2_2$, respectively.\footnote{In~\cite{Hristov:2013spa}   coordinates $({\sf x}, {\sf y})$ on the 2d base are used so that the two-dimensional metric is conformally flat, $\dd s^2_2 = \rme^{2\psi}(\dd {\sf x}^2 + \dd {\sf y}^2)$, and the volume form is ${\rm vol}_2= \rme^{2\psi}\dd {\sf x} \wedge \dd {\sf y}$. In this case one can write $A_{(2)}= \frac{1}{2} (\partial_{\sf x}\psi\, \dd {\sf y} - \partial_{\sf y} \psi\,\dd {\sf x})$, since this satisfies $\dd A_{(2)} = \frac{1}{2}  (\partial_{\sf x}^2\psi + \partial_{\sf y}^2\psi) \,\dd {\sf x} \wedge \dd {\sf y} = -\frac{1}{4}R_2 {\rm vol}_2$.} Notice that the 3d metric is the most general one that admits 
a time-like Killing vector and the gauge field is then determined in terms of the metric.

We now consider the boundary metric~\eqref{metric_bdry} together with the gauge field~\eqref{gfdexpbdy} and impose the supersymmetry conditions. This is most conveniently done using the variables introduced in section~\ref{sec:newvars} and imposing $m={g}/{\alpha}={\bt}/{(\alpha^3c)}$ together with the expression for $c$ in \eqref{cbps} and the one for $\alpha$ in \eqref{alphabps}. 
The following analysis focuses on the conformal boundary of the bulk solution and we will
not impose the extremality condition. Thus, the analysis applies to the 
conformal boundary for the class of supersymmetric bulk solutions for which, when all parameters are real, 
only the extremal case does not have a naked singularity.

Next we make the coordinate transformation
\begin{align}\label{coord_trans_canonical}
t\,=\, \kappa(1+\bt^2)\,\tilde t\,,\qquad\qquad 
\phi\,=\,\tilde\phi+\alpha(b+\bt)\,\tilde t\,,
\end{align}
so that the supersymmetric Killing vector, given in \eqref{susyvector_Lorentzian}, takes the form \eqref{K_canonical}. Then matching the metric with \eqref{3d_susy_background} we find
\be
\Upsilon^2\,=\,\frac{\left[Z(1-x^2)+2\mu(\bt^2+1)x\right]^2}{4\mu^2}\,,
\ee
and
\begin{align}
\tilde\omega & \, =\, \left(1-x^2\right) \left[4 \alpha  \left(b^2+1\right) \mu ^2 \Upsilon^2\right]^{-1}\times\nn
&\big[4 \mu ^2 (b+s) (b s-1) \left(s^2 x^2+1\right)+4 b \mu 
   \left(s^2+1\right) Zx  +b \left(1-x^2\right) Z^2\big]\dd\tilde\phi\,.
\end{align}
Here $x=\cos\theta$ as before, and to slightly simplify the formula we denoted
\be
Z \, \equiv \,  1-2 b \bt-\bt^2\,.
\ee
We also find that the two-dimensional metric reads 
\be
\dd s^2_2 \,=\, g^{(2)}_{xx}\,\dd x^2 + g^{(2)}_{\tilde\phi\tilde\phi}\,\dd \tilde\phi^2\,,
\ee 
with 
\begin{align}
 g^{(2)}_{xx}  & \, =  \, \left(b^2+1\right) \left(s^2+1\right) \left(s^2 x^4+1\right)^2 \left(Z^2-4 \mu ^2 (b
   s-1)^2\right)\times \nonumber\\
&  \qquad \Big\{\Upsilon ^{2} \big[\left(s x^2 (b s-1)+b+s\right)^2+\Upsilon
   ^2\big] \big[(x-1) Z-2 \mu  (s x (b+s)-b s+1)\big] \nonumber\\
&  \qquad\qquad \times\big[2 \mu  (s x (b+s)+b s-1)+(x+1) Z\big]\left(x^2-1\right)\Big\}^{-1}\,,
\end{align}
and
\be
g^{(2)}_{xx}\, g^{(2)}_{\tilde\phi\tilde\phi} \,=\, \frac{(\bt^2 + 1)^2(\bt^2x^4+1)^2}{\Upsilon^6}\,.
\ee
It follows that
\begin{align}
\vol_2 
\,&=\, -\frac{ \left(\bt^2+1\right) \left(\bt^2 x^4+1\right)}{|\Upsilon|^3}\,\dd x\wedge \dd\tilde\phi\,,
\end{align}
the minus sign being due to the fact that the positive orientation induced on the two-dimensional base is given by $\sin\theta \,\dd\theta\wedge\dd\tilde\phi=-\dd x\wedge\dd\tilde\phi$.

For the boundary gauge field \eqref{gfdexpbdy}, we would like to extract $f$ and $A_{(2)}$ and check the differential relations \eqref{eq_omega}.
In the coordinates given in \eqref{coord_trans_canonical} we can write
\be
A_{\rm bdy} \, = \, A_{\tilde t}\,\dd\tilde t+A_{\tilde\phi} \dd\tilde\phi\,,
\ee
with
\begin{align}
A_{\tilde t} \,=\, \kappa (1+\bt^2)A_t+\alpha(b+\bt)A_\phi  \,,\qquad
A_{\tilde\phi} \,=\, A_\phi\,,
\end{align}
where $A_t,A_\phi$ can be read from \eqref{gfdexpbdy}.
After imposing supersymmetry we obtain
\begin{align}
A_{\tilde t} &\, = \, 
\frac{ Z \left[\bt \left(1-2 b \bt-b^2\right) x^2 +b (b \bt-2)-\bt\right] x}{2 \alpha \mu \left(b^2+1\right)   \left(\bt^2 x^4+1\right)} \,,\nn[1mm]
A_{\tilde\phi}&\,  = \, \frac{2\mu Z\left(\bt^2+1\right)\left[1-b\bt+\bt(b+\bt)x^2\right]x}{\left(\bt^2x^4+1\right)\left[Z^2-4\mu^2(b\bt-1)^2\right]}\,.
\end{align}
Since the gauge potential needs to match the canonical form only up to a gauge transformation, we also allow for a shift 
\be\label{kappears}
A_{\rm bdy}\,  \to \, A_{\rm bdy} + k\, \dd \tilde t \,,
\ee
with $k$ a constant. Matching the resulting expression with~\eqref{3d_susy_background} gives
\begin{align}
f&\, = \, -A_{\tilde t}-k \, = \,  -\frac{ Z \left[\bt \left(1-2 b \bt-b^2\right) x^2 +b (b \bt-2)-\bt\right] x}{2 \alpha \mu \left(b^2+1\right)   \left(\bt^2 x^4+1\right)}-k\,,
\end{align}
and
\begin{align}
A_{(2)}\,&=\,  \, \left(s^2+1\right) \Big\{ 4 \mu ^2 \Upsilon ^2\left(b^2+1\right) Zx \left(s x^2 (b+s)-b s+1\right)\nn[1mm]
& + \Big( x Z
   \left(s x^2 \left(b^2+2 b s-1\right)+b (2-b s)+s\right)-2k \alpha\mu   \left(b^2+1\right)  
   \left(s^2 x^4+1\right)\Big)\times \nn[1mm]
&   \left(1-x^2\right) \Big(4 \mu ^2 (b+s) (1-b s) \left(s^2
   x^2+1\right)-4 b \mu Z \left(s^2+1\right) x +b Z^2\left(x^2-1\right) \Big) \Big\}  \nn[1mm]
 &\quad \  \Big[ 2\mu\Upsilon^2\left(b^2+1\right) \left(s^2 x^4+1\right)
   \left(Z^2-4 \mu ^2 (b s-1)^2\right) \Big]^{-1} \dd\tilde\phi\,.
\end{align}

We have checked that with the above ingredients both equations in \eqref{eq_omega} are satisfied, provided that we make the gauge choice
\eqref{kappears} with $k$ given by
\begin{align}
k\, = \, \frac{(b+\bt) (1-b \bt)}{\alpha  \left(b^2+1\right)}\,.
\end{align}


\providecommand{\href}[2]{#2}\begingroup\raggedright\endgroup

\end{document}